\documentclass[preprint,review,12pt]{elsarticle}

\usepackage{amssymb}
\usepackage{amsmath}
\usepackage{bm}
\usepackage{tabularx}
\usepackage{multirow}
\usepackage{nomencl}
\makenomenclature

\usepackage[misc,clock,geometry]{ifsym}

\usepackage{lineno}
\modulolinenumbers[5]
\usepackage{hyperref}

\usepackage{xcolor}

\usepackage{subcaption}
\usepackage[labelformat=parens,labelsep=quad,skip=3pt]{caption}

\journal{}

\begin{document}

\begin{frontmatter}



\title{\textbf{Mitigation of fine hydrophobic liquid aerosols by polydispersed uncharged and charged water droplets}}

\author[IITD]{Debabrat Biswal}
\author[IITD]{Bahni Ray\corref{mycorrespondingauthor}}
\author[IITD]{Debabrata Dasgupta}
\author[IITB]{Rochish M. Thaokar}
\author[IITB]{Y.S. Mayya}

\address[IITD]{Department of Mechanical Engineering, Indian Institute of Technology Delhi, Hauz Khas, New Delhi - 110016, India}
\address[IITB]{Department of Chemical Engineering, Indian Institute of Technology Bombay, Powai, Maharashtra - 400076, India}
\cortext[mycorrespondingauthor]{Corresponding authors:\\bray@iitd.ac.in(Bahni Ray), Tel: (91)-11-2659-6393, Fax: (91)-11-2658-2053}

\begin{abstract}
One of the harmful contaminants in the atmosphere, which negatively affects the well-being of both humans and animals, is the suspended respirable particles. The most difficult aspect of the study is now removing these fine respirable particles from the atmosphere. This study investigates the scavenging phenomenon of fine hydrophobic liquid aerosols (10 nm to 1050 nm) by uncharged and charged droplets in a self-made scaled test rig. In this study, a hollow cone nozzle with a 1 mm orifice diameter uses tap water to disperse liquid into fine droplets. The paraffin oil and Di-Ethyl-Hexyl-Sebacat (DEHS) solution are aerosolized to be scavenged by water droplets. This research employs a high-speed imaging technique and theoretical modeling approach to measure the size distribution and charge acquired by water droplets respectively. The findings of this study show that uncharged droplets dispersed through a hollow cone nozzle exhibit a scavenging efficiency of approximately 55\%. On the other hand, charged droplets demonstrate significantly higher scavenging efficiency, about 99\%. The investigation also reveals that the scavenging efficiency of both uncharged and charged droplets at lower potentials (1 and 2 kV) is nearly identical.
\end{abstract}



\end{frontmatter}
\newpage
\section{\label{introduction}Introduction} 
One of the primary concerns for both human health and the environment is the release of particulate matter carried in flue gases emitted from industrial and automotive exhausts \cite{carotenuto2010wet}. Among all the pollutants, the most detrimental to air quality are the very fine aerosols found in solid, liquid, and gaseous forms. In recent years, the concentration of these fine ($\le$ 2 $\mu m$) and ultrafine ($\le$ 0.2 $\mu m$) \cite{di2015capture} particles in the atmosphere has surpassed that of micron-sized particles \cite{mohan2008comprehensive}. These tiny particles can directly infiltrate the bloodstream and human lungs. Prolonged exposure to unhealthy environments leads to various incurable ailments, such as cardiovascular and pulmonary diseases \cite{wang2019influence}.
Due to the adverse catalytic effect of such pollutants on the environment, researchers and scientists are facing a challenging task in controlling and reducing the concentration of fine and ultra-fine particles in the atmosphere. This challenge has driven them to devise various mechanisms and equipment to tackle the issue effectively \cite{jaworek2006wet, zhao2008modeling, jaworek2013submicron}. \\
Several conventional air filtration methods have been developed to eliminate these hazardous particles from the environment. The techniques such as electrostatic precipitators (ESP) \cite{kim1999experimental, leonard1982experimental}, cyclones \cite{lin2013high}, wet scrubbers \cite{zhao2008modeling, mohan2008comprehensive, borra2018review,mohan2008performance, yang2010removal}, and fibrous filters \cite{lee1982theoretical, steffens2007collection} have been employed for this purpose. Additionally, HEPA (High-Efficiency Particulate Air) filters are commonly used to capture tiny particles. Through various mechanisms, such as inertial impaction, inertial interception, and Brownian diffusion \cite{ladino2011experimental,zhao2008modeling}, these tiny particles are effectively removed in the HEPA filter. Nevertheless, with extended use, the pores of the HEPA filter become blocked, resulting in a substantial drop in pressure. This increase in pressure ultimately leads to a reduction in the filter's effectiveness over time, necessitating its eventual replacement \cite{payet1992penetration}. The Stokes number and Reynolds number are crucial in influencing inertial impaction, while the Peclet number and Reynolds number affect the diffusion mechanism. The conventional mechanisms, such as cyclone separators, and wet scrubbers, are effective for larger particle sizes, typically exceeding 10 $\mu$m.
Electrically driven devices like electrostatic precipitators, electro-spray-based air purifiers, and charged droplets have been developed to remove the fine particulate matter smaller than 2 $\mu$m. The functioning of electrostatic scrubbers and electro spray is based on the basic principle of Coulomb attraction between charged droplets and particles. The particle removal mechanism using electrospray and ESP has been the subject of numerous experimental and numerical studies. These studies are delineated in the next section.\\

T. Gary et al. \cite{tepper2007electrospray}, conducted research on an electrospray technique that utilizes zero pressure drop and ozone for the removal of small aerosols through electrospray wick sources. The study revealed that this electrospray-based purification method performed effectively for a wide range of particle sizes while consuming less water and power. The efficiency of air purification is influenced by particle size, air flow rate, and system design factors. 
W. Balachandran et al. \cite{balachandran2003efficiency, jaworek2006multi}, performed experiments to remove uncharged cigarette smoke and charged smoke obtained through a corona charger using fine droplets produced by a rotary atomizer. The results showed that charged droplets significantly enhanced the removal efficiency compared to uncharged droplets. Specifically, the removal efficiency was found to be four times higher for charged smoke particles compared to uncharged smoke.
An experimental investigation was carried out to assess the particle removal efficiency of ESP and electrospray by K. Jung–Ho et al. \cite{kim2010electrospray}. The combined performance of electrospray and ESP yielded improved results in particle collection. Furthermore, it was observed that the utilization of ESP in conjunction with electrospray consumed less energy. W. Tessum et al. \cite{tessum2014factors} investigated the influence of surfactant in the spray, particle charge, and diameter on the scavenging efficiency. The results showed that larger particles (with a diameter exceeding 2 $\mu$m) were more efficiently removed compared to smaller ones. Furthermore, employing a charged spray with surfactant (opposite polarity to the particle charge) led to the significant removal of highly charged particles. S. Singh et al. \cite{singh2021scavenging}, experimentally studied the removal of three different types of particles i.e., smoke, NaCl, and metallic aerosol using a cloud of charged droplets generated by an electrohydrodynamic atomizer. The investigation revealed that the scavenging efficiency for smoke particles and NaCl varied from 19\% to 90\% respectively within the size range of 30 to 200 nm. Additionally, metallic particles exhibited significant scavenging effectiveness for sizes larger than 30 nm. A. Jaworek et al. \cite{jaworek2002numerical} studied the trajectory and deposition of small particles by charged droplets. It was observed that uncharged droplets were ineffective for removing particles smaller than  10 $\mu$m. However, charged droplets demonstrated efficient capturing of such particles, with smaller droplets exhibiting notably high collection efficiency. An investigation was conducted on the removal of particulate matter emitted from diesel engines using an electrostatic water spraying scrubber by H. H. Tran et al. \cite{ha2010enhancement}. The study revealed that as the engine's load increased, the concentration of particulate matter also increased, and the removal efficiency by uncharged spray also improved. Additionally, it was observed that the scavenging efficiency of particles by charged droplets increased by 4 to 7 times compared to uncharged droplets under constant engine load conditions. S. Lipeng et al. \cite{su2019effects}, experimentally examined how particle morphology and hydrophilicity influenced the scavenging efficiency of particles by uncharged and charged droplets. Uncharged droplets achieved higher collection efficiency for nearly spherical particles in comparison to agglomerated particles, but this efficiency decreased when the droplets were charged. Moreover, charging the droplets improved the removal efficiency of hydrophobic particles.\\

The removal of particles by ESP and electrospray has been the subject of numerous experimental and numerical research. The majority of the existing literature has studied the removal of bigger-sized particles. However, it is rare to find research that concentrates on scavenging nanometre-sized particles. According to the literature, in order to remove nanometre-sized particles, a higher voltage should be used. This may cause electric breakdown and the atmosphere may be harmed by the creation of ozone. There should be the optimum working parameter (applied potential, spray parameter, and selection of aerosol) to make an effective electrospray-based air cleaner for fine particles.\\ 
To preserve these factors for future generations, the present experimental work shows how charged and uncharged droplets remove hydrophobic liquid aerosol in the nanometre size range. This research also emphasizes determining the removal efficiency of aerosol along with the water consumption required for this study.

\section{\label{experimental setup}Experimental setup and methodology}
\subsection{Experimental setup}
\begin{figure}[h]
\centering
\includegraphics[width=1\linewidth]{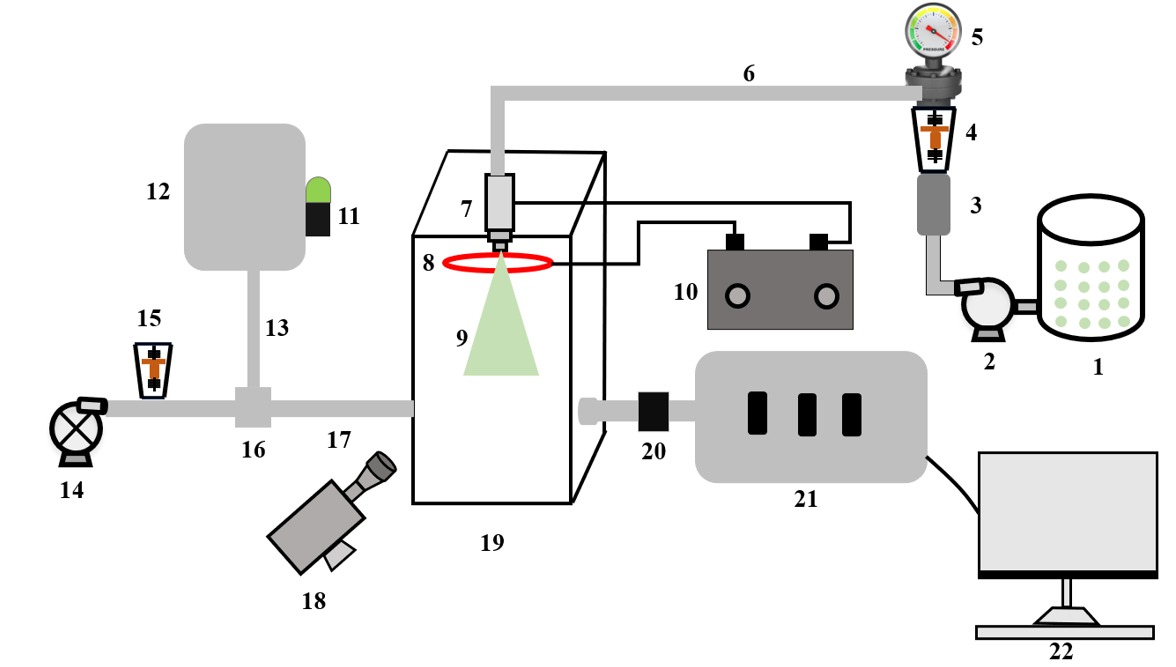}
\caption{ Lay-out of experimental arrangement for scavenging of fine aerosols by uncharged and charged droplets: (1) water reservoir (200 litter), (2) water pump, (3) strainer, (4) rota-meter for water, (5) pressure gauge, (6) water line pipe (8 mm silicon tube), (7) nozzle, (8) copper ring electrode, (9) spray, (10) high power DC source, (11) solutions (DEHS and paraffin oil), (12) atomizer, (13) aerosol passing tube, (14) air blower, (15) air rota-meter, (16) t- junction, (17) aerosol + compressed air line, (18) high - speed camera, (19) experimental chamber, (20) diffusion dryer, (21) SMPS, (22) computer with data acquisition system.}
\label{expt.setup}
\end{figure}

The layout of the experimental apparatus used to carry out this study is depicted in Figure \ref{expt.setup}. This experimental configuration comprises five distinct sections, namely: (1) An experimental chamber designed for particle droplet interaction, (2) Aerosol generation and feeding setup, (3) A mechanism for creating sprays, (4) The droplet charging system, and (5) A sampling unit for measuring particle size and number distribution.\\
We use a transparent 5 mm thick acrylic sheet to make the experimental chamber for droplet and particle interaction. The scavenging chamber has dimension of 600 mm $\times$ 200 mm $\times$ 200 mm (height $\times$ length $\times$ width). The volume of the chamber is a maximum of 24 liters. A spraying hollow cone nozzle with a 1 mm orifice is mounted on the top of the experimental chamber. A copper ring with a 40 mm diameter is placed below the nozzle at a distance of 10 mm separation. The nozzle and copper ring are fixed so that the center of the nozzle and ring coincide with each other.  The cross-flow of aerosols relative to the droplet falling direction occurs into and out of the chamber through 8 mm vents. The 8 mm vents are provided on two opposite side walls of the chamber. The 8 mm vents are located at a distance of 500 mm from the top of the chamber. To avoid the accumulation of water coming from the spray, a 20 mm drainage passage is provided at the bottom of the chamber.\\
The paraffin oil and Di-Ethyl-Hexyl-Sebacat (DEHS) solution (Sigmaaldrich) are aerosolized using the Topas ATM 228 atomizer. The aerosol concentration can be varied by changing the operating vapor pressure of the aerosol generator i.e., 0-800 hPa. As the generated aerosol possesses low pressure and low velocity, an external blower (BOSCH GBL620 Air Blower) with variable flow rate supplies pressurized air to feed the aerosol into the experimental chamber. The flow rate of compressed air is measured using a 10 liter/min air rotameter. \\
\subsection{Methodology}
Aerosol-air-laden flows into the experimental chamber through an inlet. After supplying the aerosol to the experimental chamber, a period of around 30 minutes is kept to maintain the steady-state flow of the aerosol. once the aerosol comes to a steady state, the initial concentration and size distribution of the aerosol are measured using the GRIMM Scanning Mobility Particle Sizer (SMPS) with Condensation Particle Counter (CPC) Model 5416.\\
Water droplets are dispersed from the top of the experimental chamber through a hollow conical nozzle on the aerosol. Here, water is pumped from a water reservoir to the hollow cone nozzle using a Crompton self-priming pump (Model mini strong I). The water flow rate and pressure are measured to be 0.6 lpm and 2.5 bar respectively, using a rotameter (maximum reading of the rotameter is 10 lpm) and pressure gauge (maximum reading is 8 bar). After a continuous spray,  a 30-minute time period is similarly maintained to get a steady state of particle removal. Here to avoid the entrapment of water droplets into the SMPS, a diffusion dryer is kept at the inlet of SMPS. Then the final aerosol concentration after spray impinging on it is measured. using SMPS.\\
To study the effect of the electric field in the spray on aerosol removal, the droplet is charged. To charge the water droplets, the ring electrode with a diameter of 40 mm is grounded, and the spray nozzle (hollow cone nozzle/needle) is connected to the positive polarity of a high-power DC source (10 kV, 10 mA, dual polarity, Ionics Pvt. Ltd). When the charged droplets impinge on the aerosol, the remaining aerosol concentration is sampled by SMPS.

\section{\label{Results} Results and discussions}
The electrostatic force of attraction and the size of the droplets have a significant role in removing fine aerosols in a wet electrostatic scrubber. This study focuses on analyzing (1) the droplet size distribution using a high-speed imaging technique, (2) the non-uniform electric field distribution and charge acquired by droplets using theoretical modeling, and (3) the effectiveness of the uncharged and charged droplets in removing aerosols.
\subsection{\label{Size distribution of droplets}Analysis of droplet size distribution using image processing technique}
\begin{figure}[h]
\centering
\includegraphics[width=0.8\linewidth]{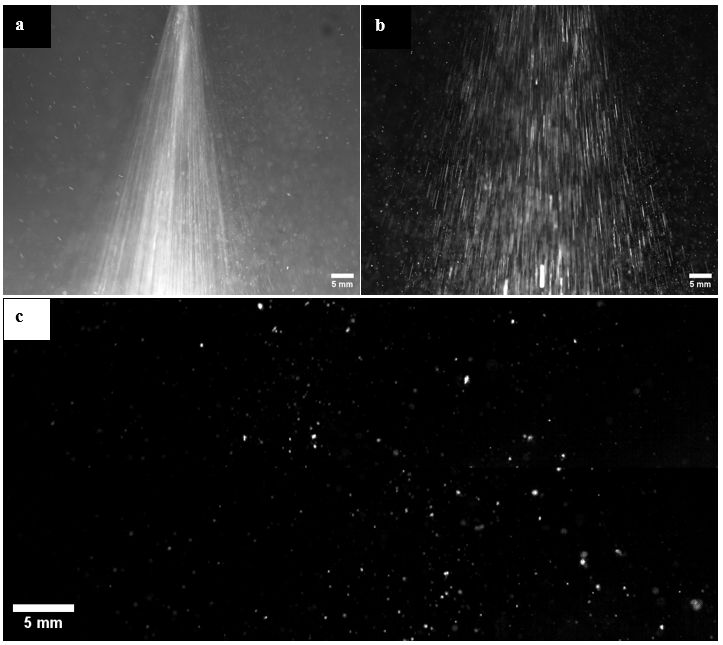}
\caption{High-speed imaging of dispersion of droplets: charged droplets at 9 kV generated through hollow cone nozzle at a flowrate of 0.6 liter/min and 2.5 bar pressure and captured at (a) 500 fps, (b) 4000 fps, (c) 32000 fps.}
\label{droplet_sizes_different_fps}
\end{figure}

\begin{figure}[h]
\centering
\includegraphics[width=0.8\linewidth]{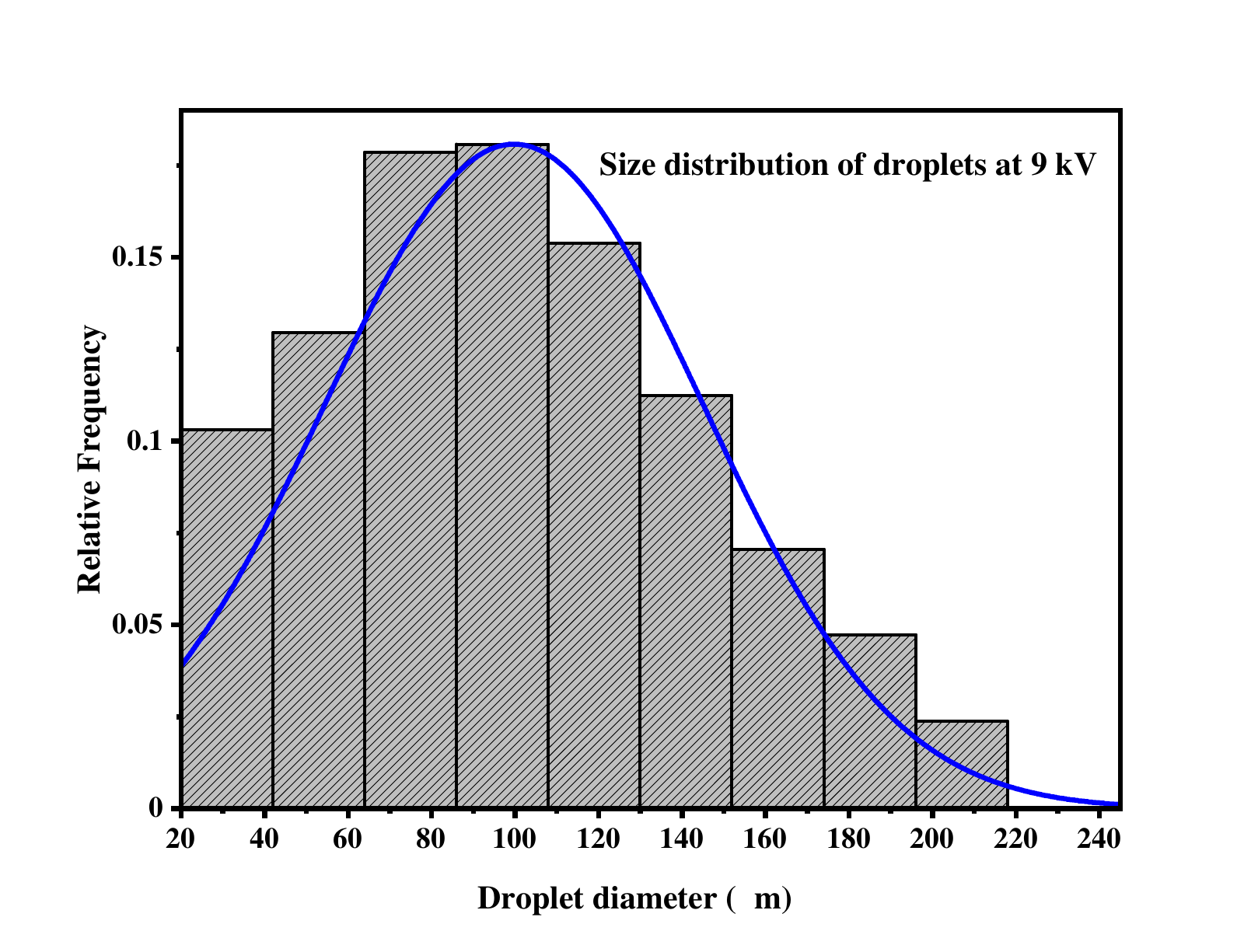}
\caption{Normal distribution of polydisperse droplets generated through hollow cone nozzle at a flow rate of 0.6 liter/min, 2.5 bar pressure, and 9 kV applied potential.}
\label{normal_distribution}
\end{figure}

 A high-speed camera (Photron MINI UX 100) connected to a high-magnification lens (Navitar 12X) is used to visualize and record the spray phenomenon to evaluate droplet size distribution. Here the spray is generated using a hollow cone nozzle with a flow rate of 0.6 liter/min and a pressure of 2.5 bar. The droplets are charged using a high-power DC source with various applied potentials from 1 kV to 9 kV with a 1 kV step-size increment. An LED light source is utilized to illuminate the spray regime. A series of spray images at different locations is visualized and captured using a high-speed camera with various fps (frames per second) as shown in Figure \ref{droplet_sizes_different_fps}. During the analysis of droplet size distribution, the following assumptions are considered.
 \begin{itemize}
     \item There is no evaporation of water droplets.
     \item No entrapment of air molecules in droplets.
     \item No coalescence occurs among the droplets.
     \item Droplets are assumed to be spherical.
 \end{itemize}
Figures \ref{droplet_sizes_different_fps} (a) and (b) show the image of the spray captured at 500 fps and 4000 fps respectively. In this image, the droplets appear distorted and spike. Therefore at lower fps (500 and 4000 fps), it is difficult to determine the size of the droplets generated through the spray. Figure \ref{droplet_sizes_different_fps} (c) represents the images of the droplets captured at a camera speed of 32000 fps. This image distinguishes the droplet and the shape of the droplet looks spherical.\\
However, the series of images captured at higher camera speed (32000 fps) are considered for droplet size distribution with the help of image processing method. Subsequently, the high frame rate (32000 fps) images are analyzed using image processing software known as "Image J (version 2.9)". The steps followed to do the droplet size analysis are discussed below.\\
The hundred images are selected from different spray regimes for image processing. Each raw image is imported to "Image J" software and converted to an 8-bit binary image. The image is subtracted from a reference image and processed for auto-thresholding. The binary image is filtered through a "fill hole" algorithm to avoid empty holes inside the droplets that appear due to illumination effects. The area of each spherical droplet is determined using the "particle analyze" algorithm with maintaining a sphericity ratio between 0.9 - 1. The roundness of an object equals 1 for a perfect spherical and  $<{\displaystyle \,}$ 1 otherwise. The equivalent diameter of each droplet is determined using equation \ref{droplet:dia} \cite{koravcin2022characterization}.
 \begin{equation}
     D_{eq}=2\sqrt{\frac{A}{\pi}}
     \label{droplet:dia}
 \end{equation}
 Here $D_{eq}$ and $A$ are the equivalent diameter of each droplet and each detected area respectively. Here we initially determine the droplet diameter in terms of pixels. To obtain the diameter of the droplets from pixels to physical units (mm or $\mu$m), the camera and lens are calibrated using a measuring scale.\\
 After the diameter of the droplets is determined and converted into $\mu$m. The size distribution of the droplets is plotted as shown in Figure \ref{normal_distribution}. It is observed that this distribution graph follows a normal distribution pattern. Figure \ref{normal_distribution} reveals that the maximum size of the charged droplets at 9 kV falls within the range of 70 – 150 $\mu$m. However, since the spray is mechanically generated using a hollow cone nozzle rather than electrically driven, the droplet size is not significantly changed at various applied potentials (1 - 9 kV).\\
 But when the spray is generated because of an external electric field, the droplet size is the function of the applied potentials. When the applied potentials are increased, the droplet size decreases. As the applied potential increases, the electric field also increases which causes enhancement of electric stress. The electric stress leads to the breakage and fragmentation of the droplets into tiny ones at higher applied potential. The high electric stress causes a reduction in droplet diameter \cite{castillo2017electrospray}.
 As from the existing collision mechanisms theory, all the collision efficiency inversely depends on the droplet diameter. Consequently, when the applied potential is increased, the droplet diameter decreases, leading to a higher collision efficiency.

\subsection{Determination of charge induced on the droplet through mathematical modeling}
To remove fine particles through the electrostatic scrubbing method, the charged spray can be generated through various conventional mechanisms, (1) through the method of induction \cite{biswal2024charged}: the charge is induced due to high voltage applied on the surface of a jet of water ejected from mechanically forced nozzle \cite{chang1987electrostatic,schmidt1992investigations}. (2) Electrospraying: The spray is produced from a capillary solely through the electrodynamic interaction of the electric field acting on the liquid's surface \cite{jaworek2006multi,wang1986charged}. (3) Corona charging: the droplets produced by the nozzle traverse through ionic current created by an electric discharge \cite{xu2003discharge}. Out of the above three mechanisms of charged spray generation, the charging of droplets through induction is employed for this experimental work.\\
After droplet size distribution is determined, the second key parameter of this study is to find out the charge of the droplets. Here the droplets get charged due to the method of induction. Different experimental studies are utilized to measure the charge of the droplets. Due to the constraints of experimental equipment in this study, a theoretical approach is adapted to evaluate the charge of the droplets. \\
A mathematical expression given in Eq. \ref{equation:charge of droplets} is employed to determine the charge of the droplets \cite{cho1964contact}. 
\begin{equation}
q = 4\pi \epsilon_0 (1.65Ea)
\label{equation:charge of droplets}
\end{equation}
Where, $\epsilon_0$ is the permittivity of free space, $E$ is the electric field, and $a$ is droplet radius.\\
Due to the non-uniformity of the electric field in the pin-ring electrode configuration, evaluating the electric field strength is not a straightforward task. The expression for the electric field distribution \cite{dall2009solution, biswal2023study} in this configuration is shown in Eqs. \ref{equation:radial}, \ref{equation:axial}.
\begin{equation}
E_y(r,y) = a V \sum_{i=0}^{\infty} \left[\frac{\xi_i (y-z_i)}{[(y-z_i)^2+r^2]^{2/3}} - \frac{\xi_i (y+z_i)}{[(y+z_i)^2+r^2]^{2/3}}\right] 
\label{equation:radial}
\end{equation}
\begin{equation}
E_r(r,y) = a V \sum_{i=0}^{\infty} \left[\frac{\xi_i}{[(y-z_i)^2+r^2]^{2/3}} - \frac{\xi_i}{[(y+z_i)^2+r^2]^{2/3}}\right]
\label{equation:axial}
\end{equation}
Here $a$ is the radius of the droplet, $V$ is the applied potential, $z_i$ and $\xi_i$ are the axial position and normalized charge of the $i^th$  image charge given by recurrent relations.
\begin{equation}
z_i = \frac{a^2}{z_0+z_{i-1}}
\label{equation:axial_position} 
\end{equation}
\begin{equation}
\xi_i = \frac{q_i}{q_0}
\label{equation:normalized_charge} 
\end{equation}
To obtain the charge values of droplets, the mathematical expressions representing the electric field along both the axial and radial directions are numerically solved in MATLAB. The estimated charges of various-size droplets at different applied potentials are given in Table \ref{Table:charge}. It is observed that with the enhancement of applied potential and droplet size, the charge of the droplet increases as shown in Table \ref{Table:charge}. 140-micron size droplet acquires a maximum charge of 0.8 pC at 9 kV applied potential.  70-micron size droplet acquires a minimum charge of 0.02 pC at an applied potential of 1 kV.
\captionof{table}{Charge acquired by droplets: V is the applied potential, q$_{70}$, q$_{100}$, q$_{120}$, and q$_{140}$ are the charge acquired by 70, 100, 120, and 140 $\mu$m size droplets.}\label{Table:charge}
\begin{tabularx}{1\textwidth} { 
  | >{\raggedright\arraybackslash}X 
  | >{\raggedright\arraybackslash}X
  | >{\raggedright\arraybackslash}X
  | >{\raggedright\arraybackslash}X
  | >{\raggedleft\arraybackslash}X | }
 \hline
V (kV) & $q_{70}$ (C)  & $q_{100}$ (C)& $q_{120}$ (C) & $q_{140}$ (C)  \\	
 \hline
 1 & $2.26\times 10^{-14}$ & $4.61\times 10^{-14}$ & $6.65\times 10^{-14}$ & $9.05\times 10^{-14}$\\
 \hline
 2 & $4.51\times 10^{-14}$ & $9.22\times 10^{-14}$ & $1.33\times 10^{-13}$ & $1.81\times 10^{-13}$\\
 \hline
 3 & $6.77\times 10^{-14}$ & $1.38\times 10^{-13}$ & $1.99\times 10^{-13}$ & $2.72\times 10^{-13}$\\
\hline
4 & $9.02\times 10^{-14}$ & $1.84\times 10^{-13}$ & $2.66\times 10^{-13}$ & $3.62\times 10^{-13}$\\
 \hline
 5 & $1.13\times 10^{-13}$ & $2.31\times 10^{-13}$ & $3.32\times 10^{-13}$ & $4.53\times 10^{-13}$\\
 \hline
 6 & $1.35\times 10^{-13}$ & $2.77\times 10^{-13}$ & $3.99\times 10^{-13}$ & $5.43\times 10^{-13}$\\
 \hline
7 & $1.58\times 10^{-13}$ & $3.23\times 10^{-13}$ & $4.65\times 10^{-13}$ & $6.34\times 10^{-13}$\\
 \hline
 8 & $1.80\times 10^{-13}$ & $3.69\times 10^{-13}$ & $5.32\times 10^{-13}$ & $7.24\times 10^{-13}$\\
 \hline
 9 & $2.03\times 10^{-13}$ & $4.15\times 10^{-13}$ & $5.98\times 10^{-13}$ & $8.15\times 10^{-13}$\\
 \hline
\end{tabularx}

 \subsection{Scavenging phenomenon of liquid aerosol by charged and uncharged droplets}
In the wet electrostatic scrubber, aerosol-laden air collides with the charged droplets when it passes from one region to another.  During the collision between the aerosol and droplets, the aerosols are captured due to inertial impaction, interception, Brownian diffusion, and electrostatic force of attraction \cite{singh2021scavenging}. Once the aerosols are attached to the droplets, the removal mechanics of the aerosol is entirely controlled by the dynamics of the droplets and electrostatic force.\\
This current research focuses on the removal efficiency of two different hydrophobic aerosols. The findings of this study are elucidated in the subsequent sections.
\subsubsection{Comparison of number and mass distribution of test aerosols}
\begin{figure}[t!]
\begin{subfigure}[t!]{0.49\textwidth}
\centering
\includegraphics[width=1.2\linewidth]{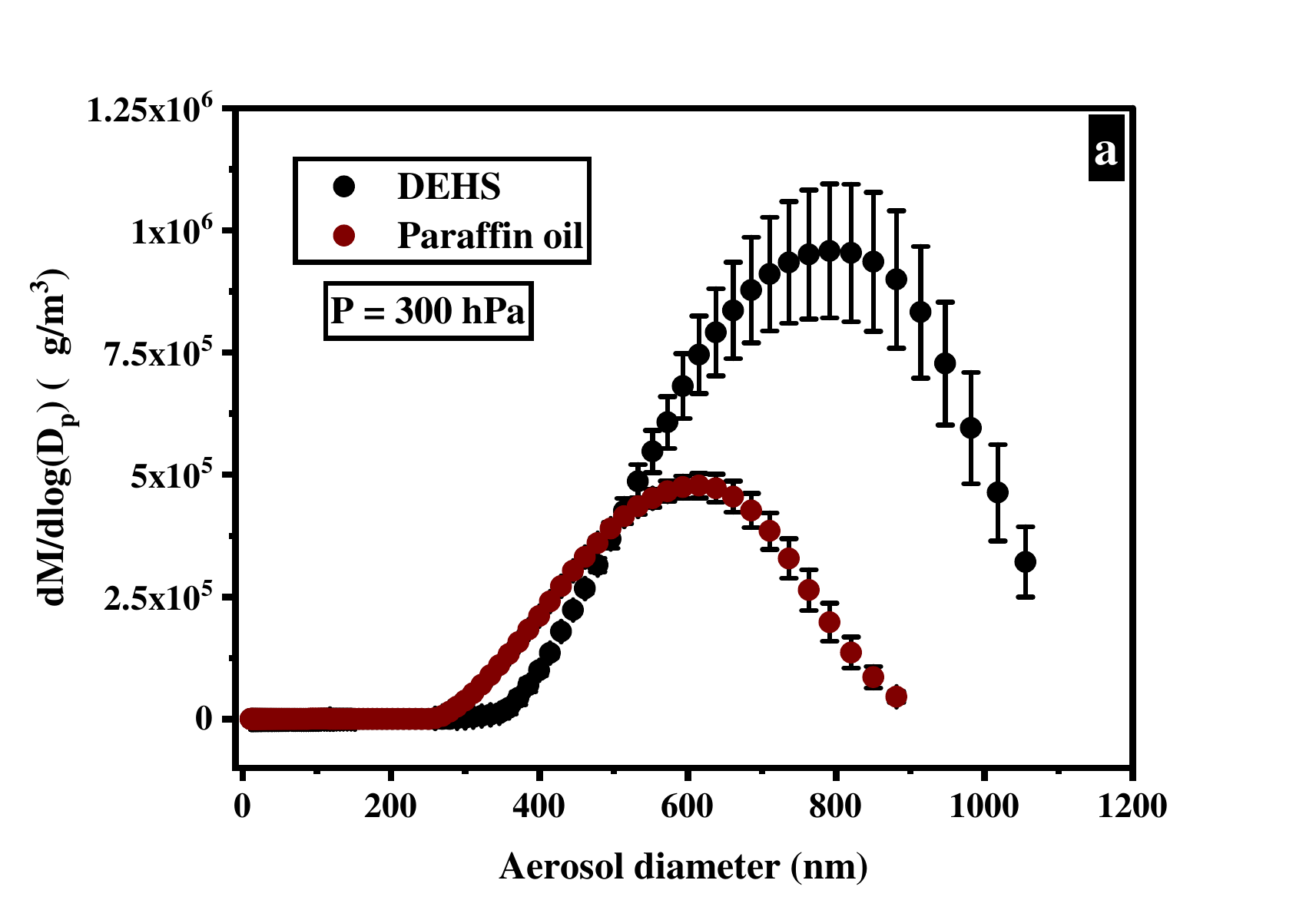}
\caption{ Mass}
\label{1.}
\end{subfigure}
\hfill
\begin{subfigure}[t!]{0.49\textwidth}
\centering
\includegraphics[width=1.2\linewidth]{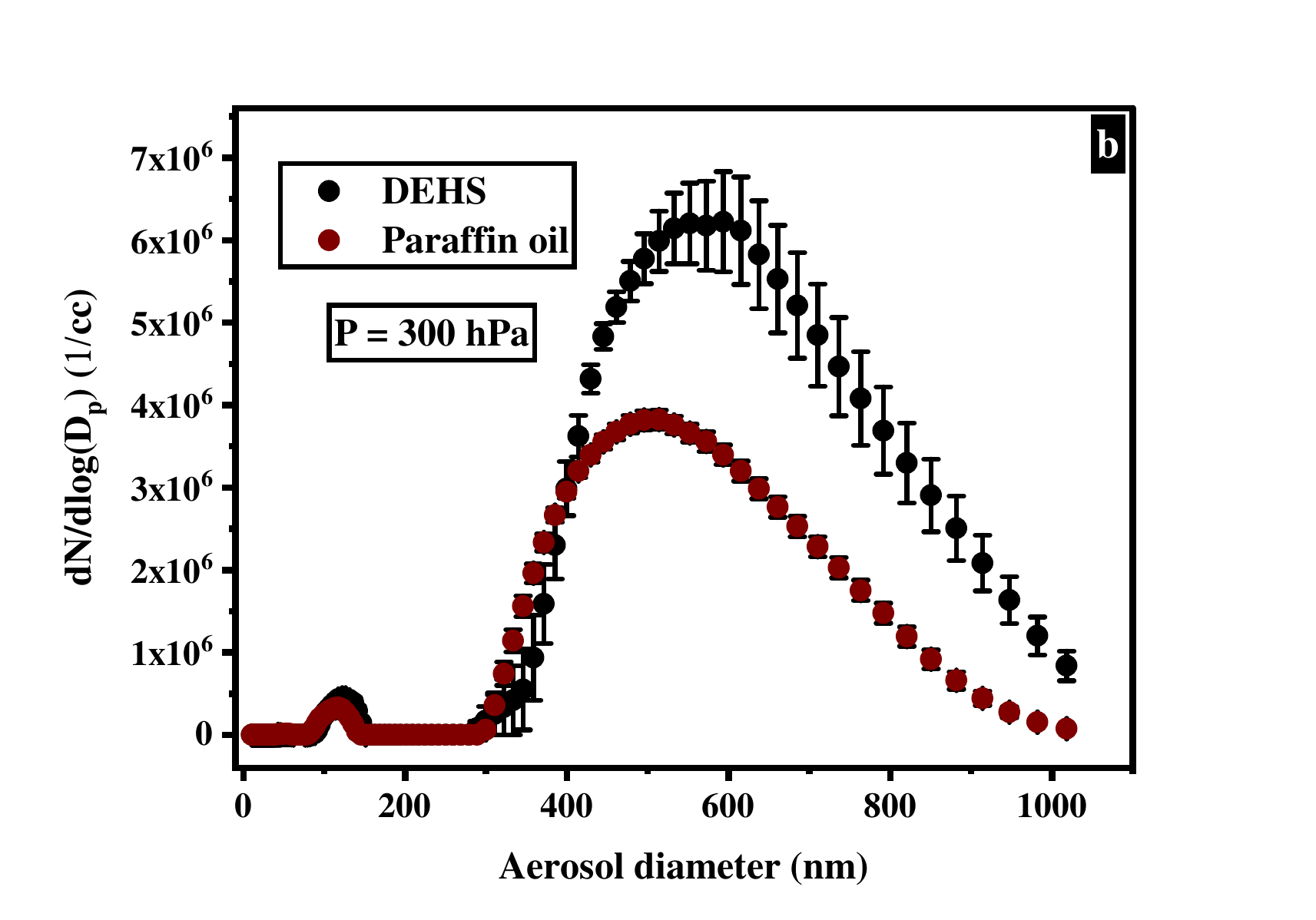}
\caption{ Number}
\label{2.}
\end{subfigure}
\caption{ Comparison of mass and number concentration of DEHS solution and paraffin oil at 300 hPa vapor pressure of atomizer without any spray.}
\end{figure}

In this study, paraffin oil and DEHS solution are atomized for fine liquid aerosol generation. These aerosols are effectively used to determine the removal efficacy of uncharged and charged droplets. Table \ref{Table:1} presents the physical and chemical properties of these liquid solutions. These solutions are hydrophobic. In this study, both solutions are compared based on the number and mass concentration of the aerosol produced by the aerosol generator. The two solutions are aerosolized at different vapor pressures of the aerosol generator. If the solution is aerosolized at high vapor pressure, the aerosol concentration will be more and vice versa. Therefore, the generation of aerosol concentration strongly depends on the operating vapor pressure of the aerosol generator. In this experiment, a series of aerosol concentrations is considered by varying vapor pressure from 100 to 500 hPa with a 50 hPa interval. Out of all the different vapor pressures, the mass and number concentration of aerosols generated from paraffin oil and DEHS solution are sampled using SMPS at 300 hPa vapor pressure.
The mass and number concentration of aerosol produced from two solutions are depicted in Figure \ref{1.} and \ref{2.}. It is observed that when the aerosol generator atomizes the DEHS solution, high concentrations of aerosol are generated as shown in Figures \ref{1.} and \ref{2.}. A comparatively lower aerosol concentration is produced when paraffin oil is considered.\\ 
When the DEHS solution is aerosolized at 300 hPa vapor pressure, the maximum number and mass concentration of aerosol produced are 2.33$\times$ $10^8$ $1/cc$ and 3.14 $\times$ $10^7$ $\mu g/m^3$ respectively. Whereas, in the case of paraffin oil, the maximum number and mass concentration of aerosol generated at the same vapor pressure were 7.44$\times$ $10^7$ $1/cc$ and 8.5$\times$ $10^6$ $\mu g/m^3$ respectively. Figure \ref{1.} shows that when DEHS solution is aerosolized, the maximum mass of the aerosol is produced in the size range of 800 - 1000 nm. When paraffin oil is aerosolized, the maximum mass of aerosols is produced in the 600 - 800 nm size range. Figure \ref{2.}  shows that the maximum number of aerosols are produced with the size range of 600 - 800 nm and 400 - 600 nm for DEHS solution and paraffin oil respectively.\\\\
This concludes that high concentrations of aerosol are generated from DEHS solution. Whereas, very fine aerosols are produced when paraffin oil is considered.
\captionof{table}{Physical and chemical properties DEHS solution and paraffin oil at 293K.}\label{Table:1}
\begin{tabularx}{1\textwidth} { 
  | >{\raggedright\arraybackslash}X 
  | >{\raggedright\arraybackslash}X  
  | >{\raggedleft\arraybackslash}X | }
 \hline
 Solution name & DEHS  & Paraffin \\
 \hline
 Color  &  White  & White \\
 \hline
 Chemical formula &  C$_{26}$H$_{50}$O$_4$  & C$_{15}$H$_{11}$ClO$_7$ \\
 \hline
 Solubility  & insoluble in water  & insoluble in water \\
\hline
Density ($kg/m^3$) &  912  & 827 - 890 \\
 \hline
 Vapor pressure (Pa)  & $ \le 1$ & 0.5 \\
 \hline
 Melting point ($^{\circ}$C) &  -48  & -24 \\
 \hline
 Surface tension (N/m) &  $3.2\times10^{-2}$ & $3.5\times10^{-2}$ \\
 \hline
  Dynamic viscosity (mPa s) &  22 - 24 & 5 - 17\\
 \hline
\end{tabularx}

\subsection{Effect of low electric field intensity on aerosol removal}
\begin{figure}[t!]
\begin{subfigure}[t!]{0.49\textwidth}
\centering
\includegraphics[width=1.2\linewidth]{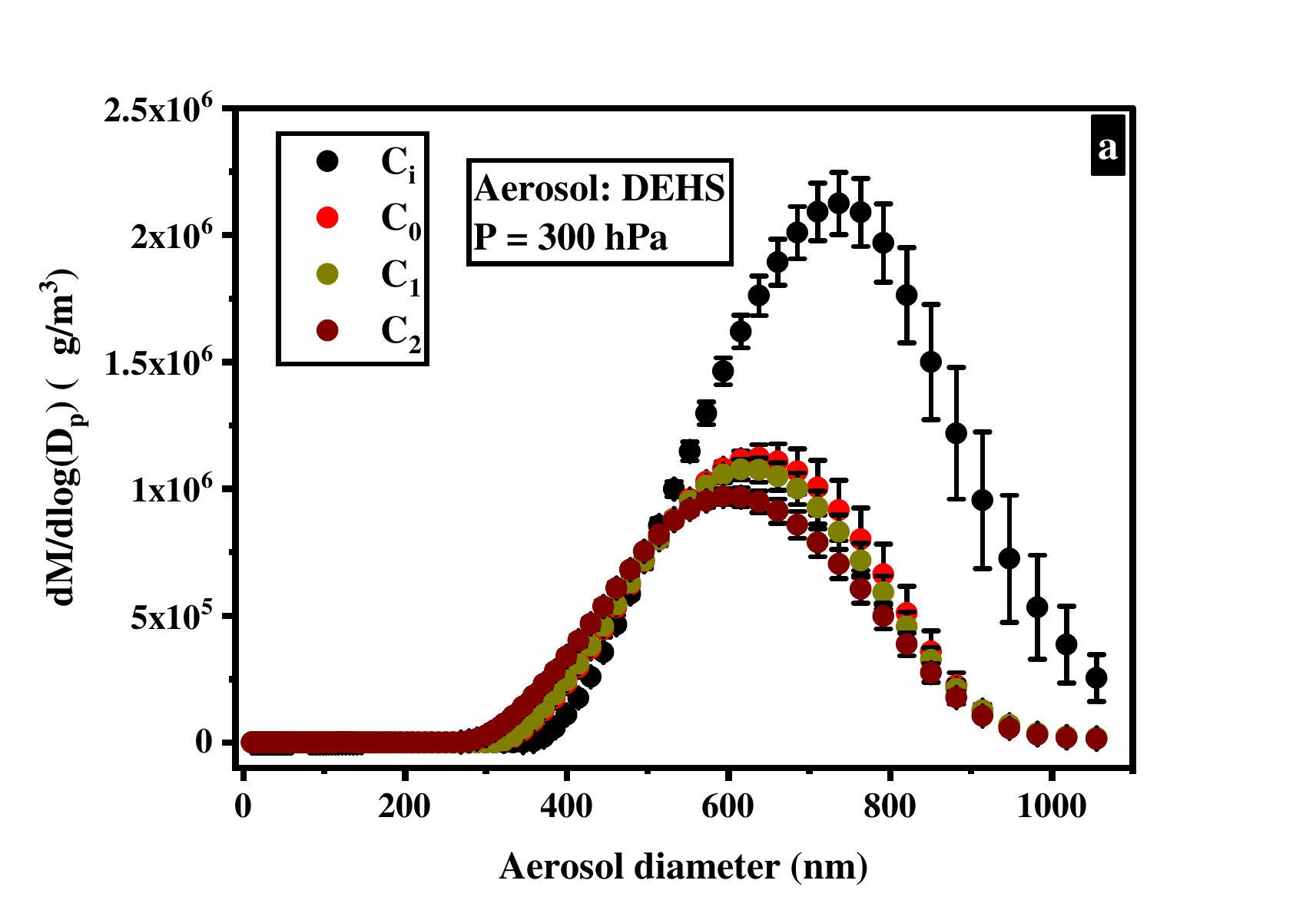}
\caption{ Mass}
\label{3.}
\end{subfigure}
\hfill
\begin{subfigure}[t!]{0.49\textwidth}
\centering
\includegraphics[width=1.2\linewidth]{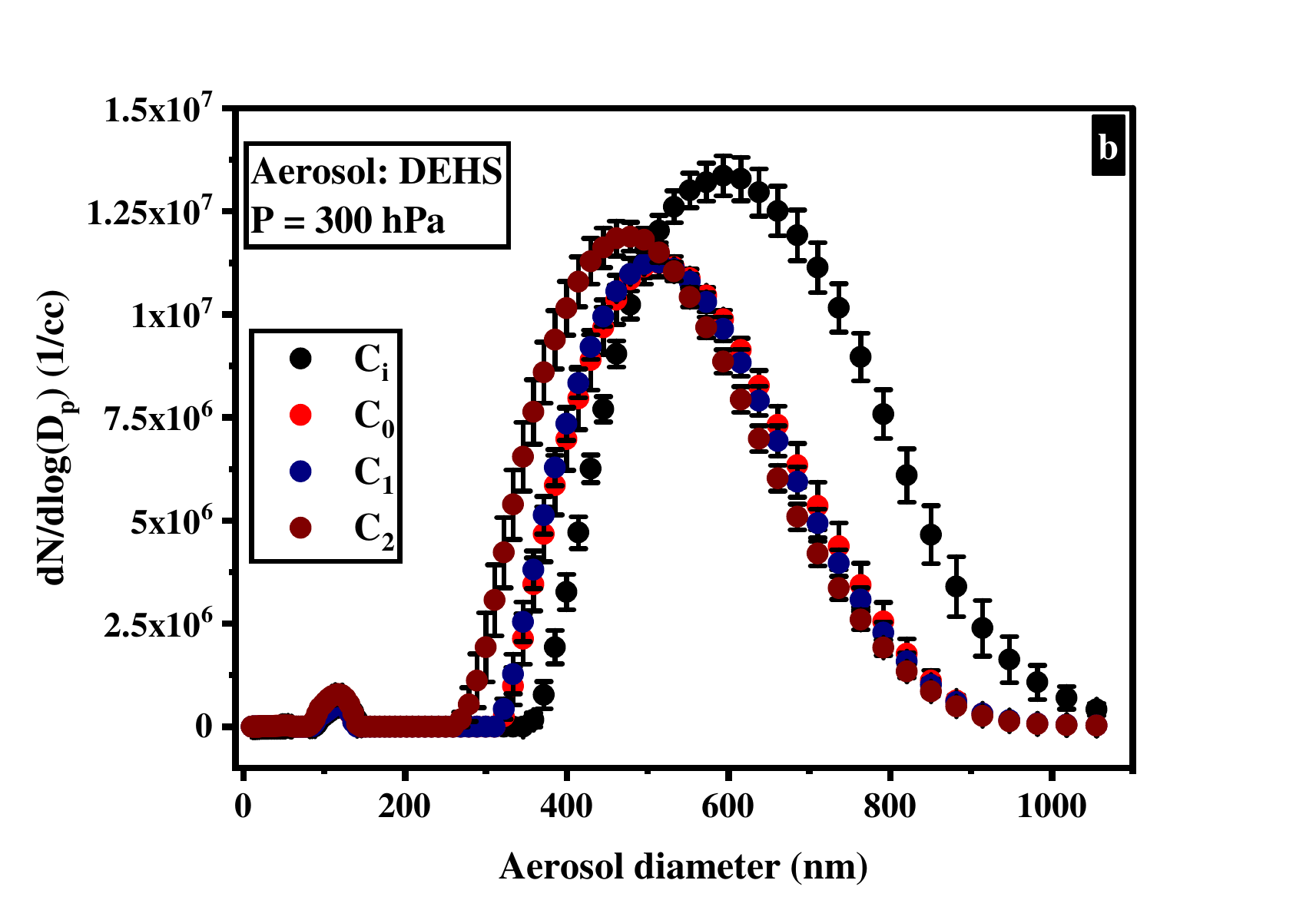}
\caption{ Number}
\label{4.}
\end{subfigure}
\caption{Change in mass and number concentration of DEHS aerosol (concentration at 300 hPa) by uncharged and charged droplets (1 and 2 kV): $C_i$: initial concentration of aerosol in the experimental chamber without any spray, $C_0$: concentration of aerosol in the experimental chamber with spray without any electric field, $C_1$ and $C_2$: concentration of aerosol in the experimental chamber with spray at applied potential of 1 and 2 kV respectively. }
\end{figure}

This section delineates the efficacy of uncharged and charged droplets on aerosol removal with a low-intensity non-uniform electric field. Here the droplets are charged at lower applied potential (1 and 2 kV). The 300 hPa vapor pressure of the aerosol generator is operated to aerosolize the DEHS solution. Before the aerosols and droplet interaction, the aerosol generated at 300 hPa vapor with compressed air is fed to the experimental chamber at a flow rate of 10 liter/min. The outlet of the experimental chamber is connected to SMPS. The sampling unit SMPS samples the aerosol at a flow rate of 0.3 L/min.\\
 The initial number and mass concentration of aerosol generated at 300 hPa are measured as 2.33$\times$ $10^8$ $1/cc$ and 3.14 $\times$ $10^7$ $\mu g/m^3$ respectively. In the first case of this work, the efficacy of uncharged droplets produced from the spray on removing fine aerosols is studied. Once the aerosol-air-laden flow becomes steady inside the experimental chamber, the spray without any external electric field impinges on the aerosols. 10 mins of the period are maintained to make a steady flow of the droplets coming from the spray. After 10 minutes of droplets and aerosol interaction, the remaining aerosol present in the chamber is sampled by SMPS. It is found that the remaining mass of DEHS aerosols present in the chamber is 1.76$\times$ $10^7$ $\mu g/m^3$ as shown in Figure \ref{3.}. This signifies that the uncharged droplets remove around half the aerosol from the experimental chamber.\\
In the next series of experiments, the droplets are charged at applied potentials of 1 kV and 2 kV. The charged droplets fall on the aerosols in the chamber. Five sets of experiments are conducted for each 1 kV and 2 kV applied potentials. It is observed that when the charged droplet at 1 kV collides with aerosols, the remaining mass of the aerosol left in the chamber is  1.7$\times$ $10^7$ $\mu g/m^3$. Similarly, the mass of the particles accumulated in the chamber after the charged droplet at 2 kV is 1.68$\times$ $10^7$ $\mu g/m^3$.\\\\
However, with uncharged and charged droplets at 1 and 2 kV applied potential, around half the aerosols are removed as shown in  Figures \ref{3.}, \ref{4.}. This indicates wet electrostatic scrubber at lower applied potentials i.e., up to 2 kV achieves the same effectiveness as a wet scrubber. This also implies that the electrostatic force of attraction has no significant role in capturing more aerosols. The above findings lead to a hypothesis that the aerosols are captured because of inertial impaction, interception, and Brownian diffusion.

\subsection{Effect of high electric field intensity on aerosol removal}
\begin{figure}[h!]
\begin{subfigure}[h!]{0.49\textwidth}
\centering
\includegraphics[width=1.2\linewidth]{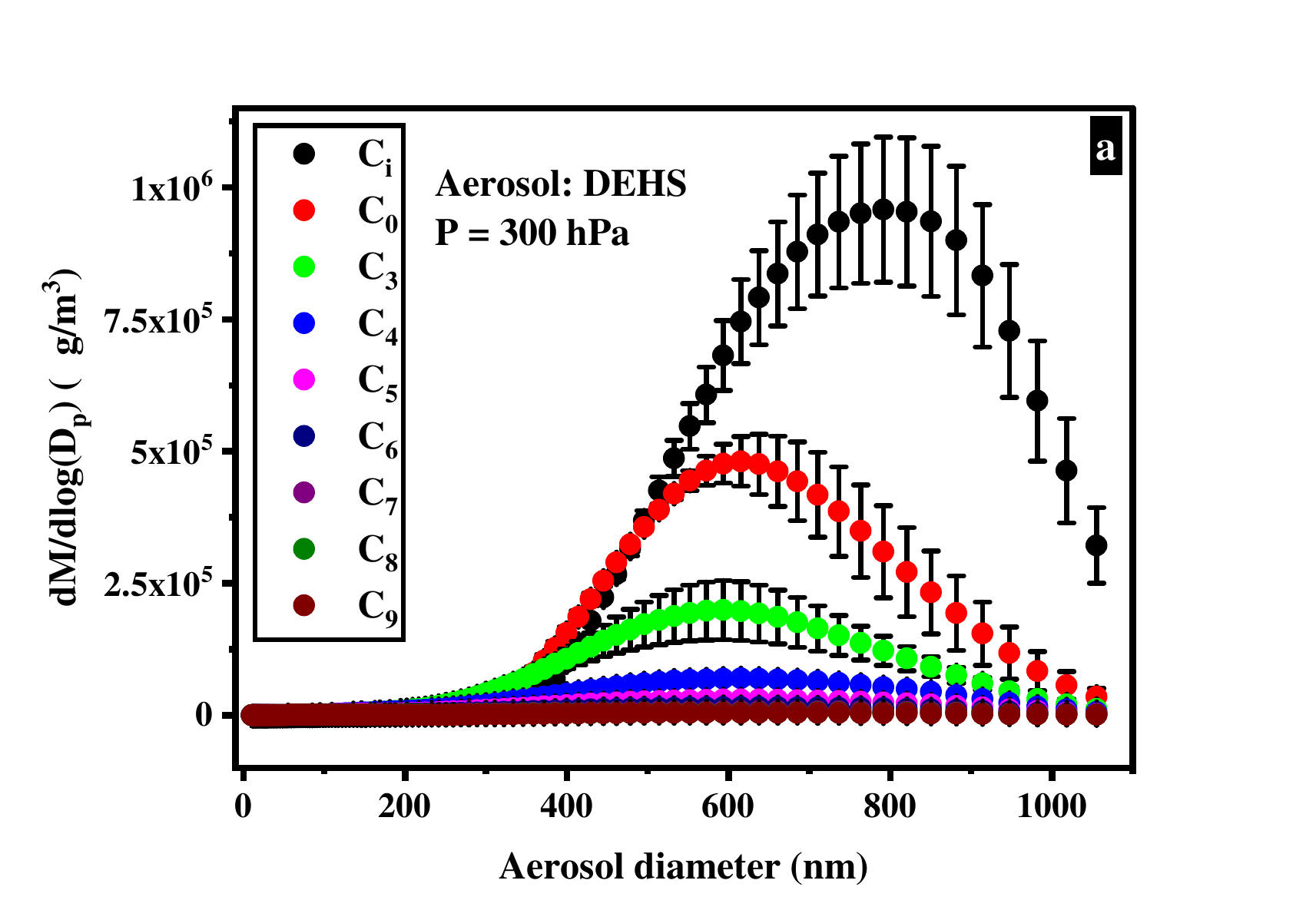}
\caption{ DEHS}
\label{5.}
\end{subfigure}
\hfill
\begin{subfigure}[h!]{0.49\textwidth}
\centering
\includegraphics[width=1.2\linewidth]{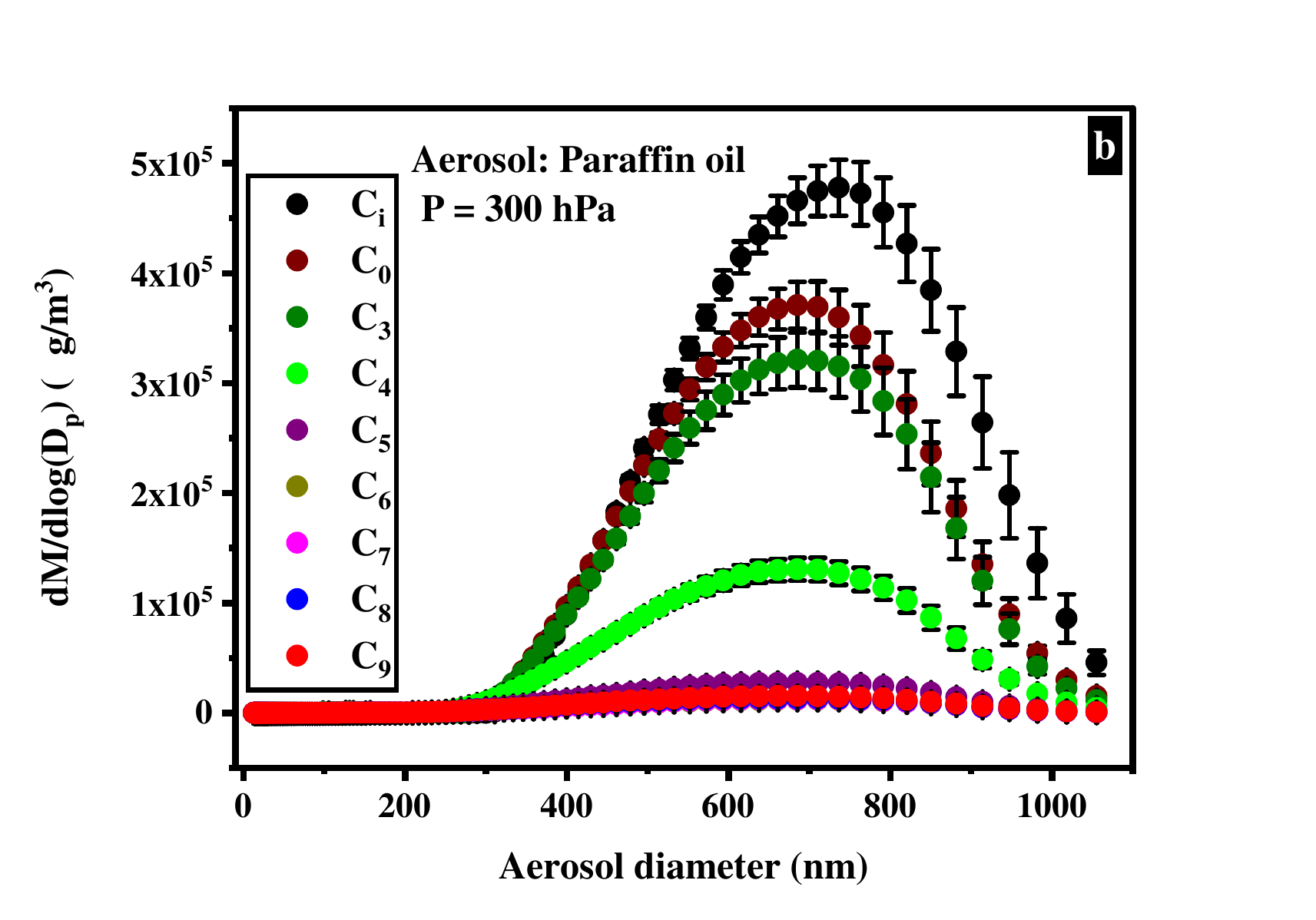}
\caption{Paraffin oil}
\label{6.}
\end{subfigure}
\caption{Change in aerosol mass concentration (concentration at 300 hPa) by uncharged and charged droplets at higher potential: $C_i$: initial concentration of aerosol in the experimental chamber without any spray, $C_0$: concentration of aerosol in the experimental chamber with spray without any electric field, $C_3$, $C_4$, $C_5$, $C_6$, $C_7$, $C_8$, and $C_9$: concentration of aerosol in the experimental chamber with charged spray at applied potential of 3, 4, 5, 6, 7, 8, and 9 kV respectively.}
\end{figure}

\begin{figure}[t!]
\begin{subfigure}[t!]{0.49\textwidth}
\centering
\includegraphics[width=1.2\linewidth]{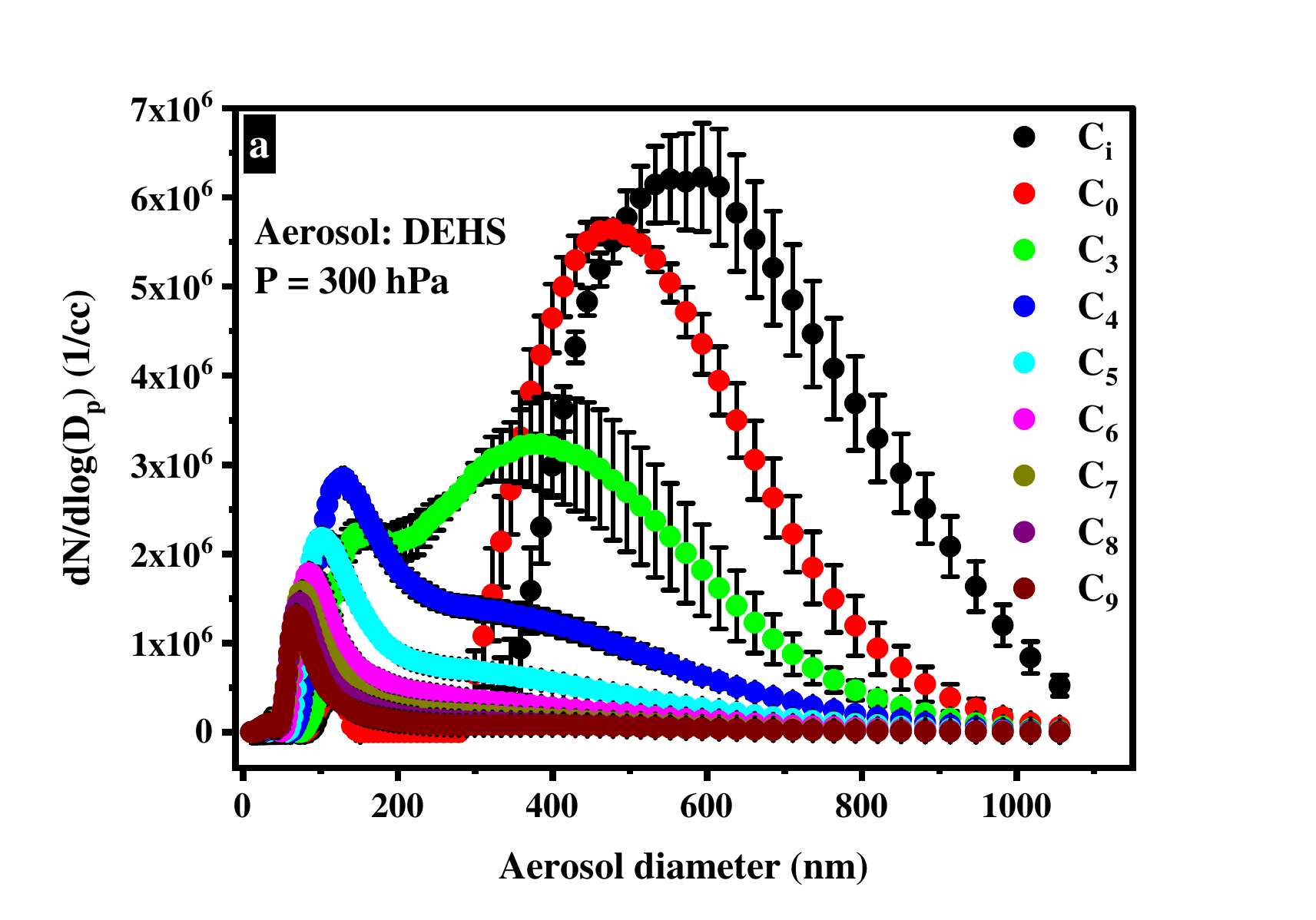}
\caption{ DEHS}
\label{7.}
\end{subfigure}
\hfill
\begin{subfigure}[t!]{0.49\textwidth}
\centering
\includegraphics[width=1.2\linewidth]{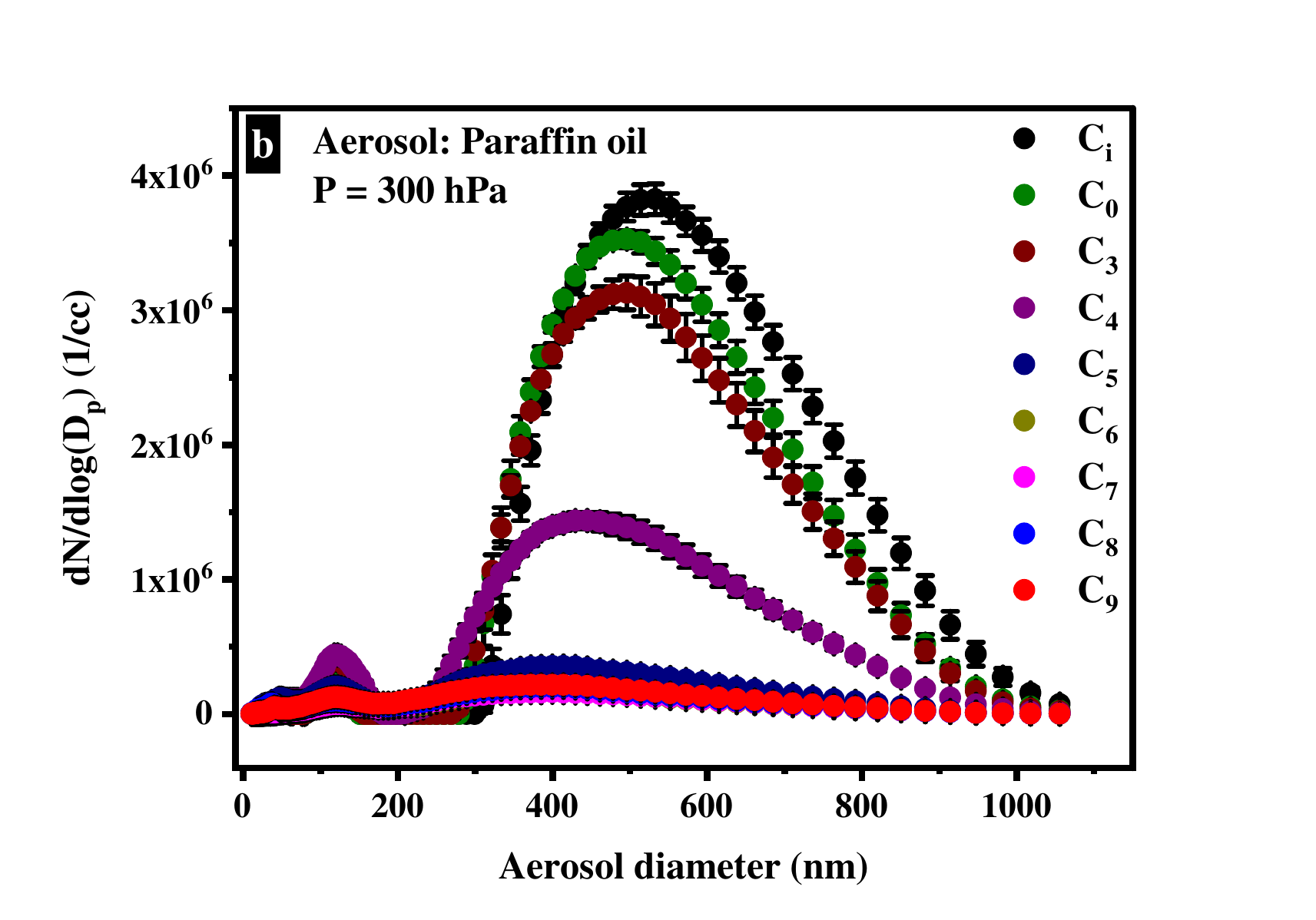}
\caption{Paraffin oil}
\label{8.}
\end{subfigure}
\caption{Change in number concentration of aerosol (concentration at 300 hPa) by uncharged and charged droplets at higher potential: $C_i$: initial concentration of aerosol in the experimental chamber without any spray, $C_0$: concentration of aerosol in the experimental chamber with spray without any electric field, $C_3$, $C_4$, $C_5$, $C_6$, $C_7$, $C_8$, and $C_9$: concentration of aerosol in the experimental chamber with spray at applied potential of 3, 4, 5, 6, 7, 8, and 9 kV respectively.}
\end{figure}

From the above-obtained results, particles removed by uncharged and charged droplets at lower applied potentials ( 1 kV and 2 kV) are not quite an emerging method for removing large amounts of the particles. The main objective of this study is to maximize the removal capacity of the charged droplet for cleaning the aerosols. This study employs a high-electric field intensity between the electrodes. The droplets are charged at higher applied potentials i.e., 3 to 9 kV. It is observed that with this nozzle-ring electrode configuration, there is no occurrence of electric breakdown at 9 kV. But the electric breakdown happens at 10 kV.\\
Similar to the previous experiment, the removal of both the aerosols i.e., paraffin oil and DEHS solution are studied using charged droplets at higher applied potential. Both Figures \ref{1.}, \ref{2.} depict that the mass and number concentration of the paraffin oil possesses a lower value compared to the DEHS solution. It is observed that when the charged droplets at higher applied potentials (3 to 9 kV) collide with aerosol present in the experimental chamber,  a significant of the particles are removed as shown in Figures. \ref{5.}, \ref{6.} \ref{7.}, and \ref{8.}.\\
Before the spray on the aerosols, the initial mass and number concentrations of the DEHS aerosol are sampled as 3.14 $\times$ $10^7$ $\mu g/m^3$  and  2.33$\times$ $10^8$ $1/cc$ respectively as shown in Figures \ref{5.} and \ref{7.}. When the charged droplet at 3 kV interacts with aerosol, the remaining aerosol left in the chamber is found to be 4.51 $\times$ $10^6$ $\mu g/m^3$ in terms of mass (as shown in Figure \ref{5.}) and 1.09$\times$ $10^8$ $1/cc$ in terms of number (as shown in Figure \ref{7.}). However, when the charged droplet at 9 kV impinged on the DEHS aerosol, the remaining mass concentration of the aerosol in the chamber is 1.22 $\times$ $10^5$ $\mu g/m^3$. Similarly for paraffin oil, before the commencement of the spray in the experimental chamber, the aerosol's initial mass concentration is 8.5 $\times$ $10^6$ $\mu g/m^3$. When charged droplets 3 kV and 9 kV make interaction with aerosol, the remaining mass concentrations of the aerosol in the chamber are sampled to be 5.95 $\times$ $10^6$ $\mu g/m^3$ and 3.47 $\times$ $10^5$ $\mu g/m^3$ respectively as shown in Figure \ref{6.}.\\\\
The finding of this study signifies that when the droplets are charged at 3 kV or more, the removal of the aerosols becomes significant. Therefore, this indicates that the electrostatic force of attraction dominates other capturing mechanisms (inertial impaction, interception, and Brownian diffusion) for removing more concentration of the particles.
From above the obtained results and graphs, it is observed that the aerosols whose size lies between 10 nm to 300 nm are non-removable. This criticizes this study and recommends conducting more studies in the future for removing very fine (10 nm to 300 nm) liquid aerosols. 
\newpage

\subsection{Analysis of total aerosol concentration}
\begin{figure}[h]
\begin{subfigure}[h]{0.49\textwidth}
\centering
\includegraphics[width=1.2\linewidth]{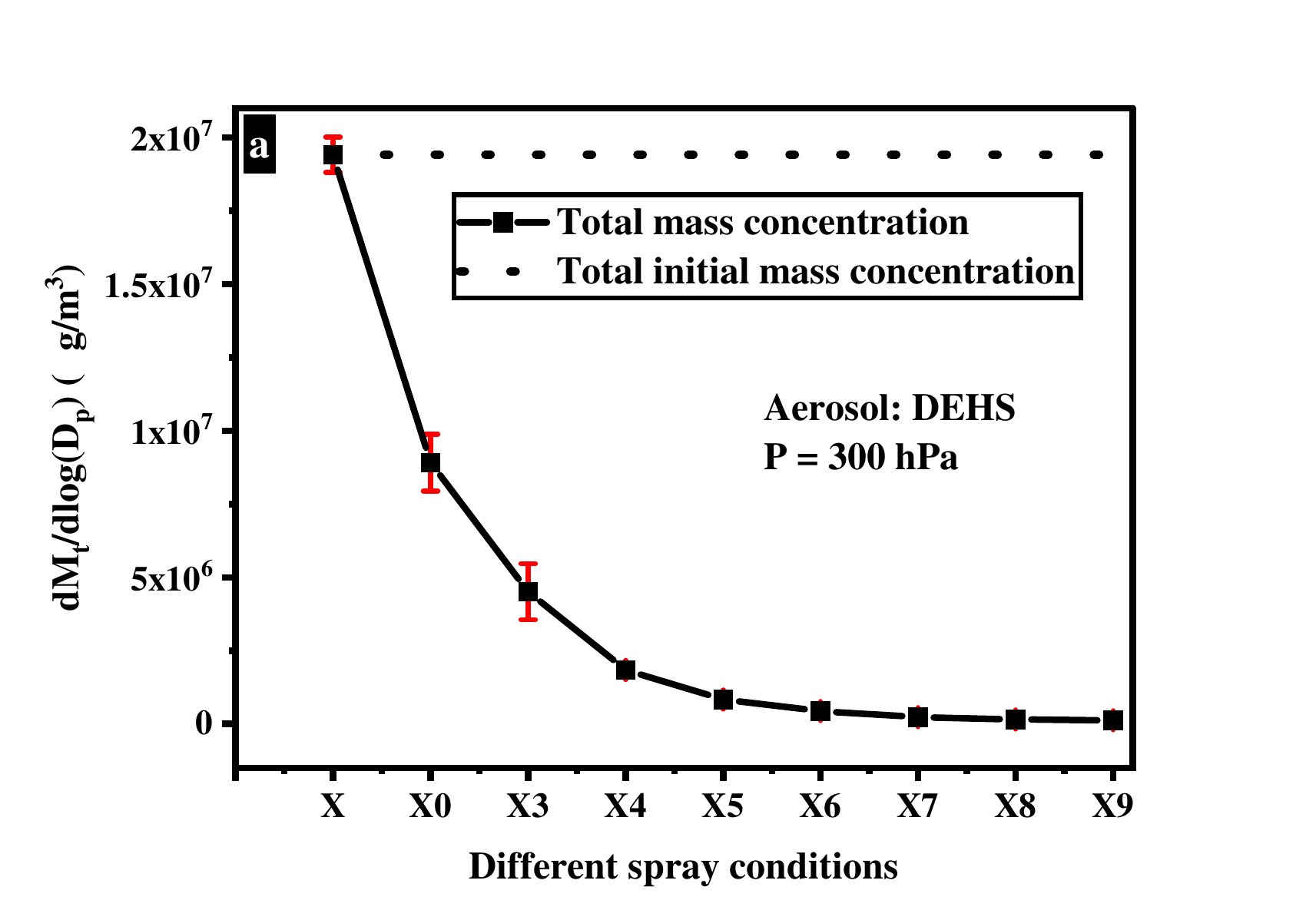}
\caption{ DEHS}
\label{9.}
\end{subfigure}
\hfill
\begin{subfigure}[h]{0.49\textwidth}
\centering
\includegraphics[width=1.2\linewidth]{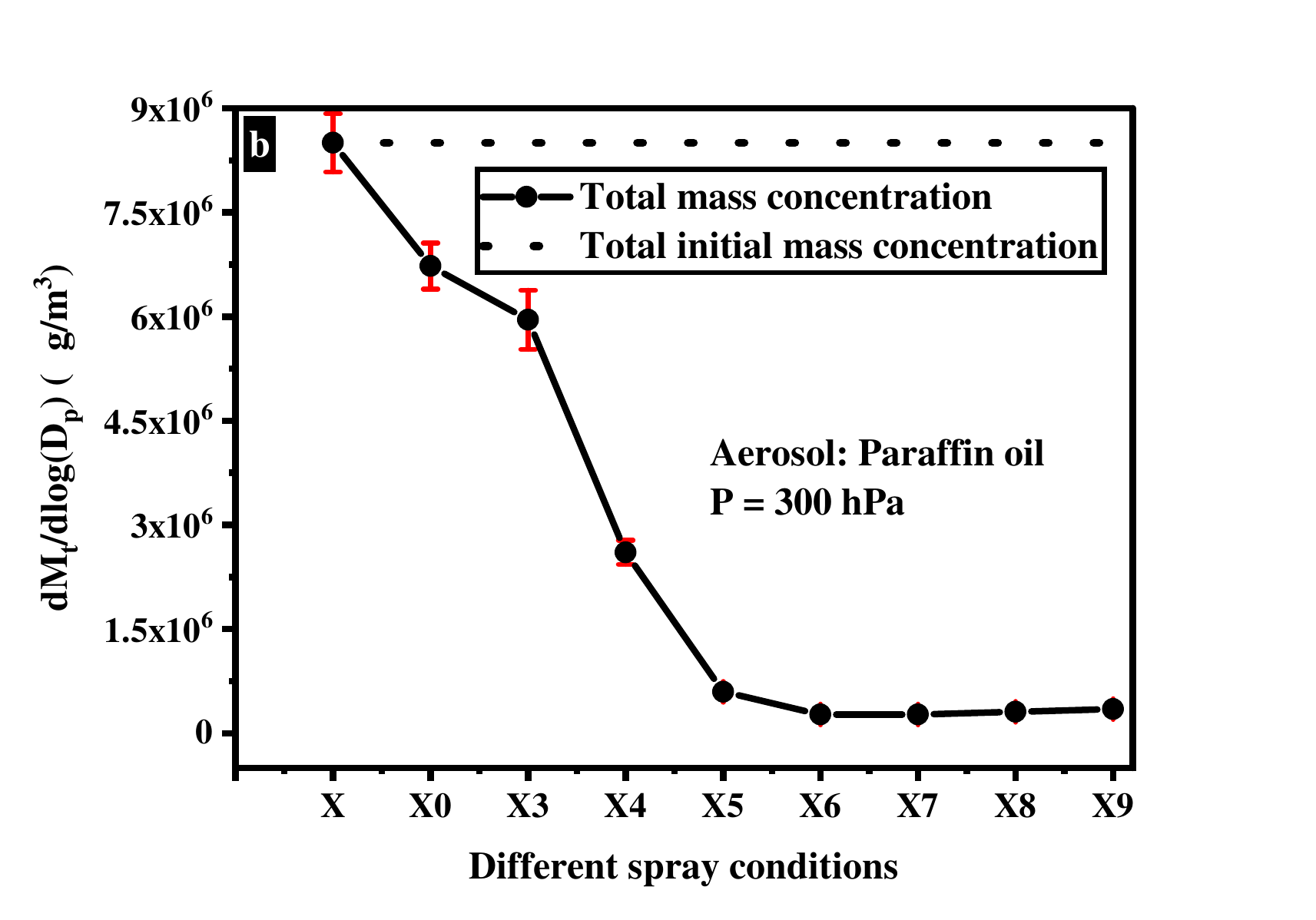}
\caption{Paraffin oil}
\label{10.}
\end{subfigure}
\caption{Change in overall mass concentration of aerosol (concentration at 300 hPa) by uncharged and charged droplets at higher potential (3 - 9 kV): $X$: no spray, $X0$: spray without electric field, $X3$, $X4$, $X5$, $X6$, $X7$, $X8$, and $X9$: charged spray at 3, 4, 5, 6, 7, 8, and 9 kV applied potentials respectively.}
\end{figure}

\begin{figure}[h]
\begin{subfigure}[h]{0.49\textwidth}
\centering
\includegraphics[width=1.2\linewidth]{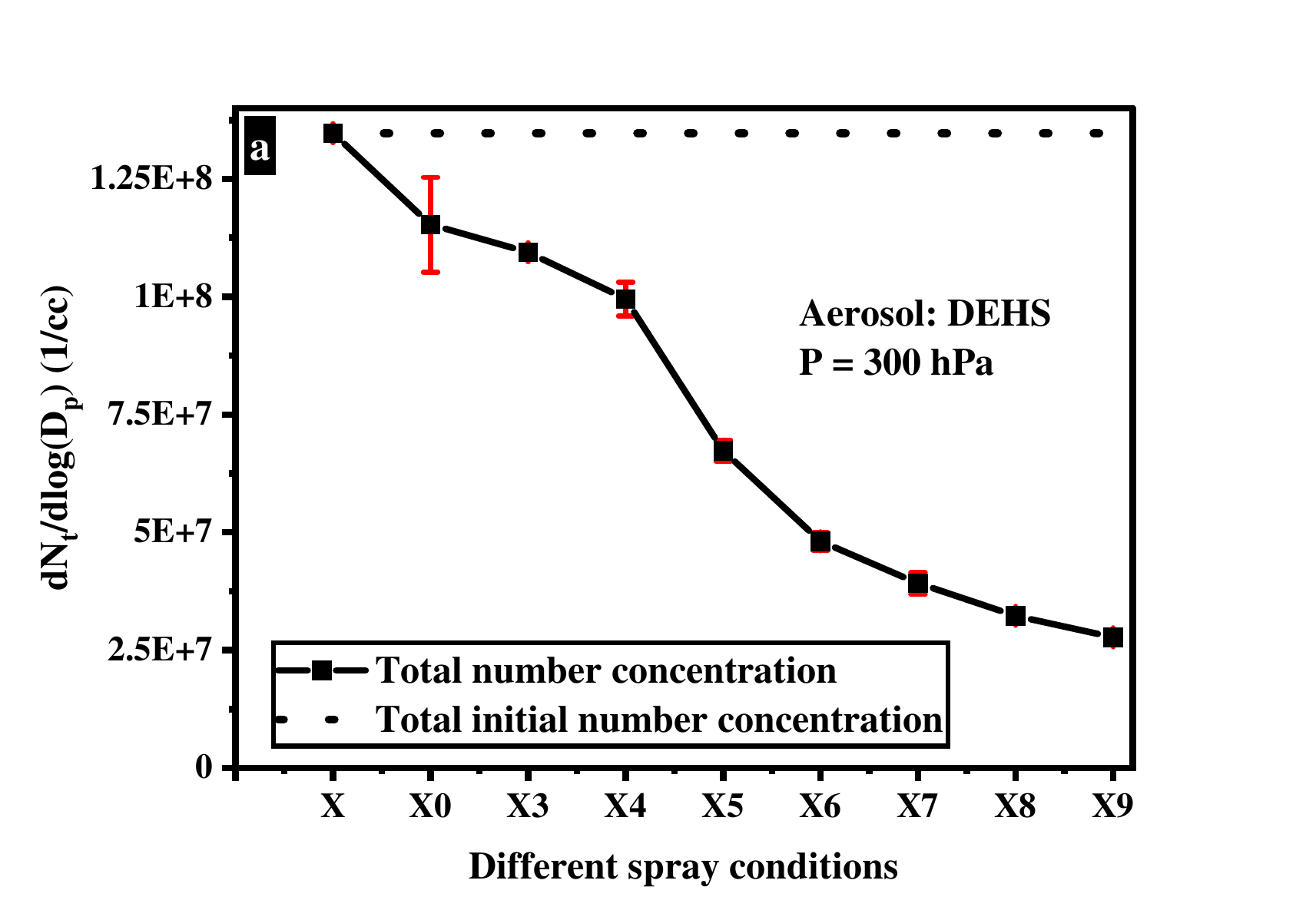}
\caption{ DEHS}
\label{15.}
\end{subfigure}
\hfill
\begin{subfigure}[h]{0.49\textwidth}
\centering
\includegraphics[width=1.2\linewidth]{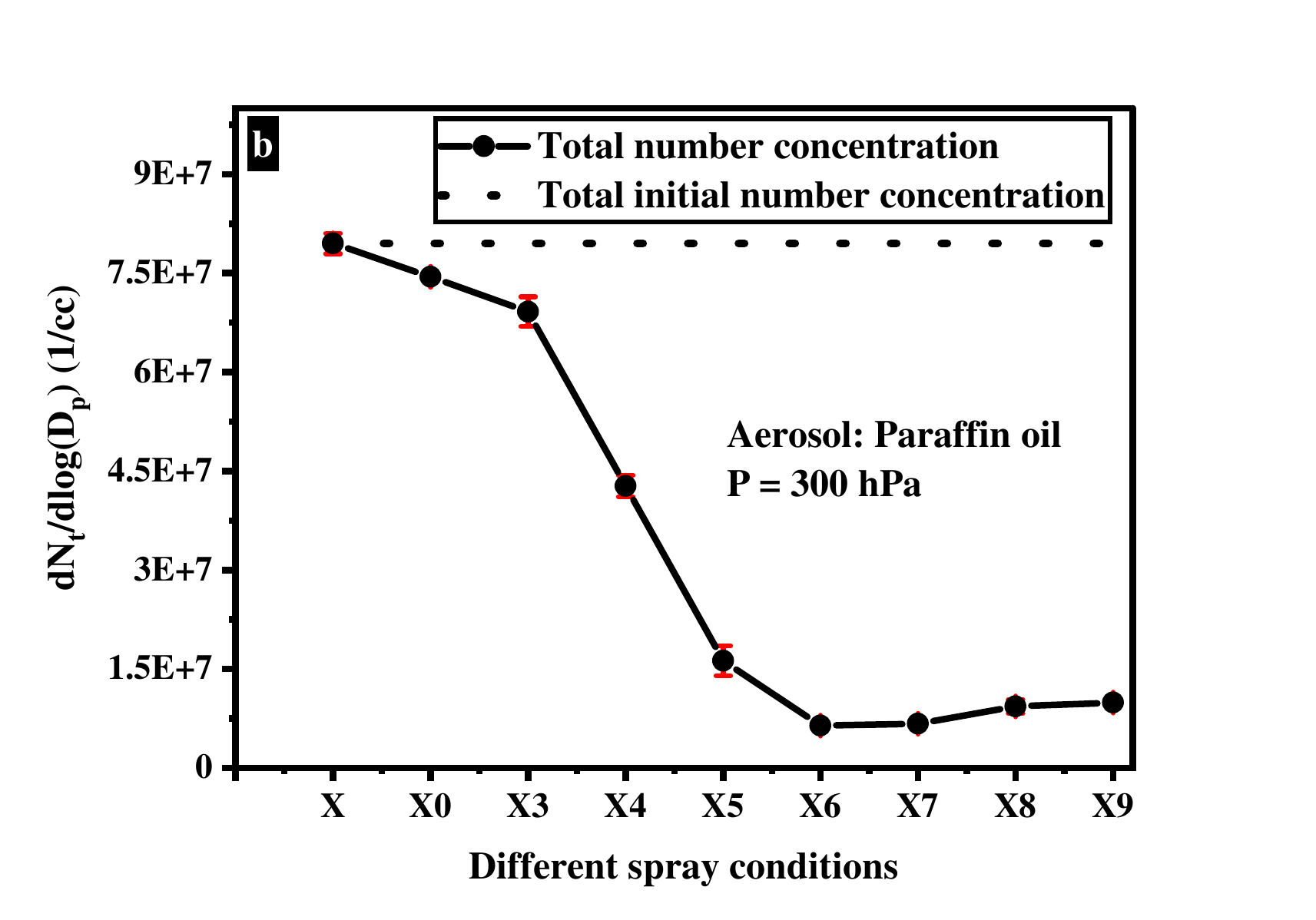}
\caption{Paraffin oil}
\label{16.}
\end{subfigure}
\caption{Change in overall number concentration of aerosol (concentration at 300 hPa) by uncharged and charged droplets at higher potential (3 - 9 kV): $X$: no spray, $X0$: spray without electric field, $X3$, $X4$, $X5$, $X6$, $X7$, $X8$, and $X9$: charged spray at 3, 4, 5, 6, 7, 8, and 9 kV applied potentials respectively.}
\end{figure}

The above findings illustrate the aerosol removals by uncharged and charged droplets considering each diameter (10 - 1050 nm) of the aerosol. This section will provide a comprehensive idea about total aerosol removal (including all the diameters) by uncharged and charged droplets. The total concentration of the aerosol is determined by adding the concentration of all the individual-size aerosols. When the DEHS aerosol is used, the initial mass concentration is $1.9\times 10^7$ $\mu g/m^3$ in the experimental chamber as shown in Figure \ref{9.}. When the charged droplets at 3 kV, 4 kV, 5 kV, 6 kV, 7 kV, 8 kV, and 9 kV applied potentials collided with aerosol, the aerosol concentration ($\mu g/m^3$.) of the chamber with corresponding applied potentials changes to $4.5\times 10^6$, $1.8\times 10^6$, $8.3\times 10^5$, $4.3\times 10^5$, $2.2\times 10^5$, $1.5\times 10^5$, and $1.2\times 10^5$ respectively as shown in Figure \ref{9.}. Similarly, the total initial number concentration of DEHS aerosol is $1.35\times 10^8$ $1/cc$. After the charged droplets interact with particles, the number concentration ($1/cc$)of the particles changes to $1.1\times 10^8$ at 3 kV, $9.9\times 10^7$ at 4 kV, $6.7\times 10^7$ at 5 kV, $4.8\times 10^7$ at 6 kV, $3.9\times 10^7$ at 7 kV, $3.2\times 10^7$ at 8 kV, and $2.7\times 10^7$ at 9 kV as shown in Figure \ref{15.}.\\
This indicates that with the enhancement of applied potentials, the removal capacity of aerosol by charged droplets gradually increases. Similarly for paraffin oil, the same trend of particle removal in terms of total number and mass concentration is observed as shown in Figures \ref{10.}, \ref{16.}.

\subsection{Analysis of scavenging efficiency of the aerosols}
Since the scavenging efficiency is the prime factor for this experimental study. The scavenging efficiency of the particles for uncharged and charged droplets is determined using mathematical expression as mentioned in Eq. \ref{equation:1} \cite{di2015capture, singh2021scavenging}.
\begin{equation}
   \eta = (1 -\frac{C_f}{C_i}) \times 100
    \label{equation:1}
\end{equation}
Here, $\eta$ is the percentage of scavenging efficiency, $C_f$ and $C_i$ are final and initial concentrations (in terms of mass or number) respectively.\\
 Since, the number concentration signifies which size of the particles can be removed using charged droplets, while the mass concentration represents the total amount of particles scavenged from the chamber. So it is necessary to determine the scavenging efficiency of the particles based on both number and mass concentrations.
 \subsubsection{Effect of aerosol size on scavenging efficiency}
\begin{figure}[t!]
\centering
\includegraphics[width=1\linewidth]{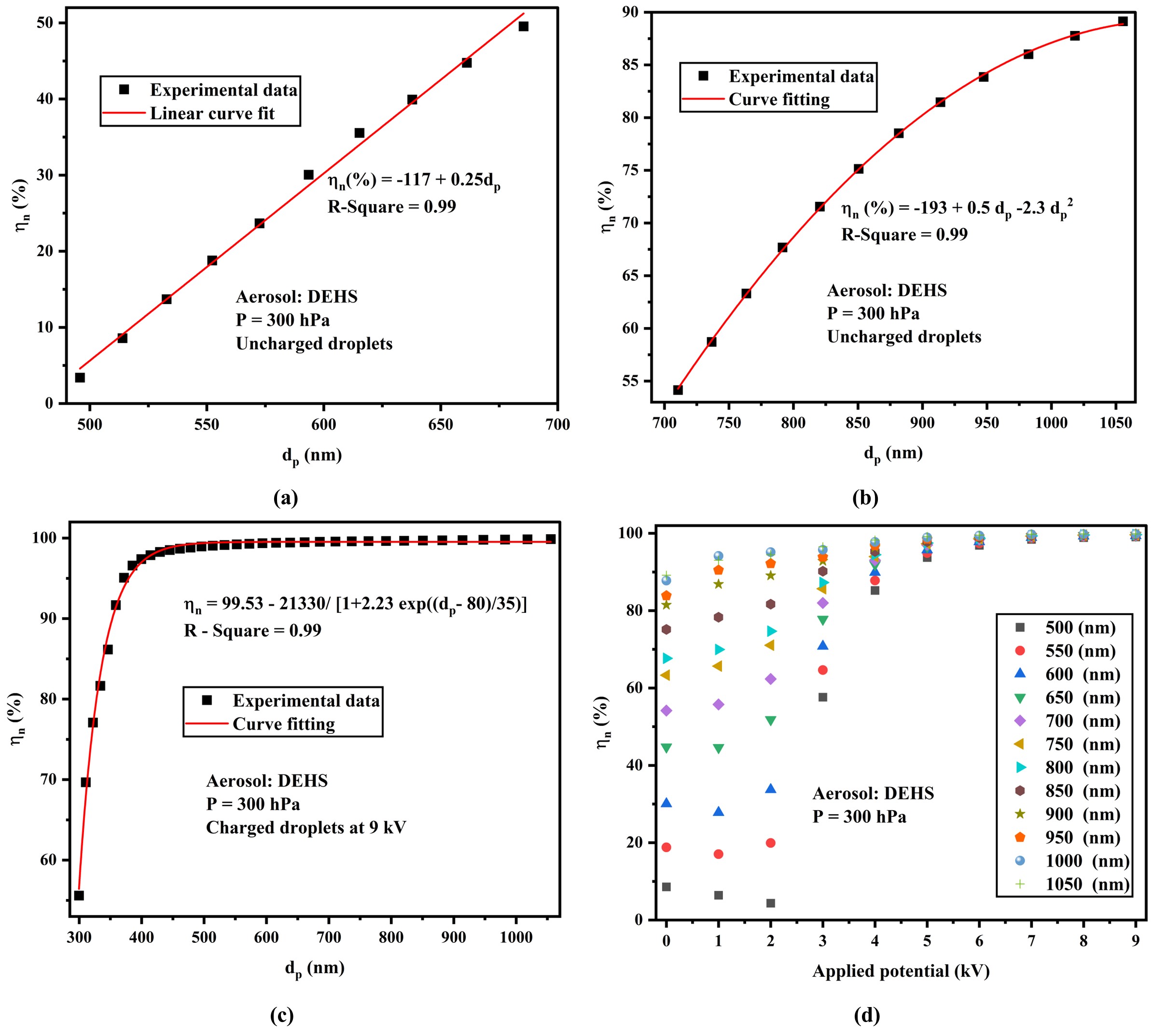}
\caption{ Scavenging efficiency based on the size of the particles of DEHS aerosol (concentration at 300 hPa) by uncharged and charged droplets: (a) aerosol size from 495 - 685 nm by uncharged droplet, (b) aerosol size from 700 - 1050 nm by uncharged droplets, (c) charged droplets at 9 kV, and (d) 500 - 1050 nm at different applied potentials.}
\label{diameter_vs_efficiency}
\end{figure}

\begin{figure}[t!]
\begin{subfigure}[t!]{0.49\textwidth}
\centering
\includegraphics[width=1.2\linewidth]{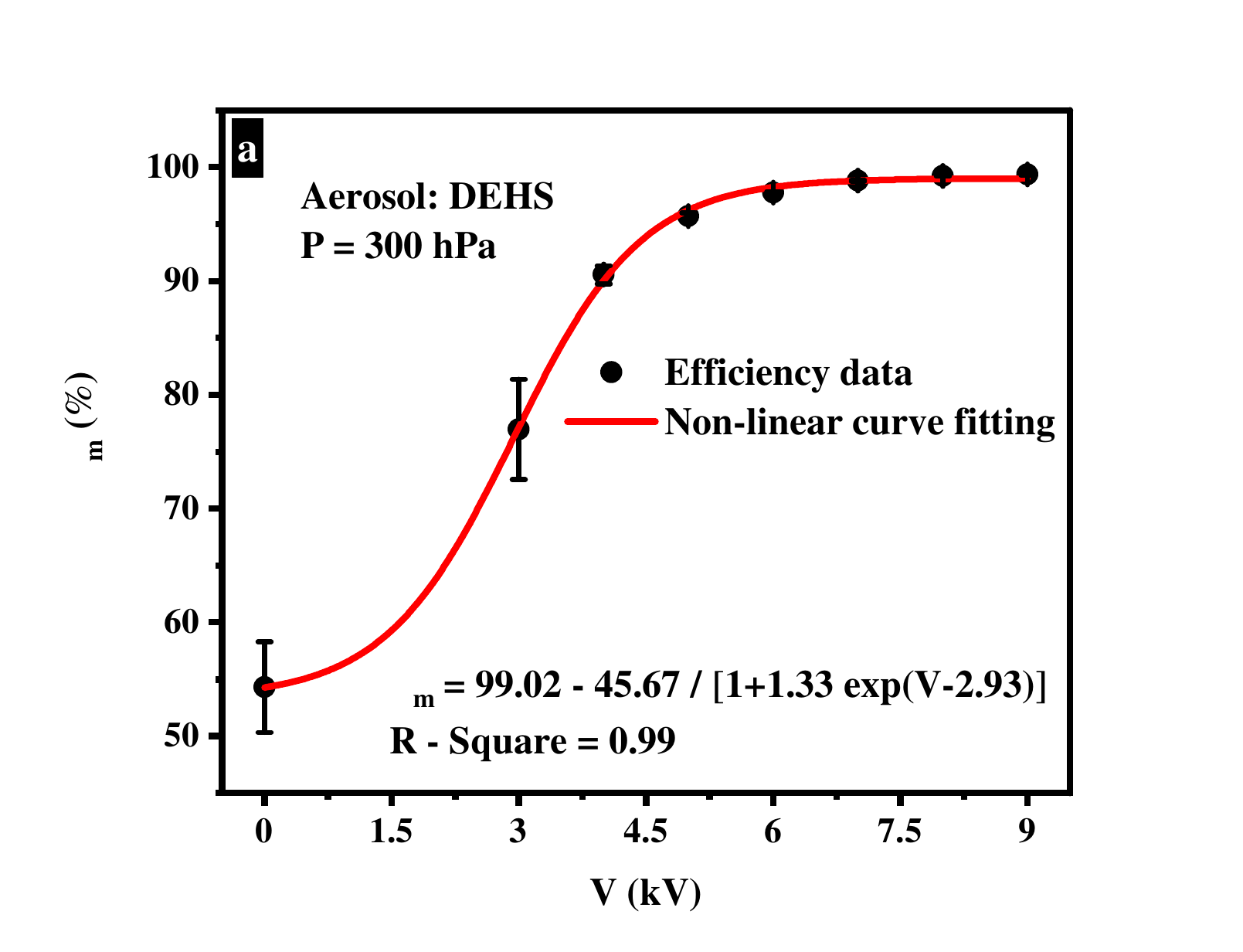}
\caption{ DEHS}
\label{11.}
\end{subfigure}
\hfill
\begin{subfigure}[t!]{0.49\textwidth}
\centering
\includegraphics[width=1.2\linewidth]{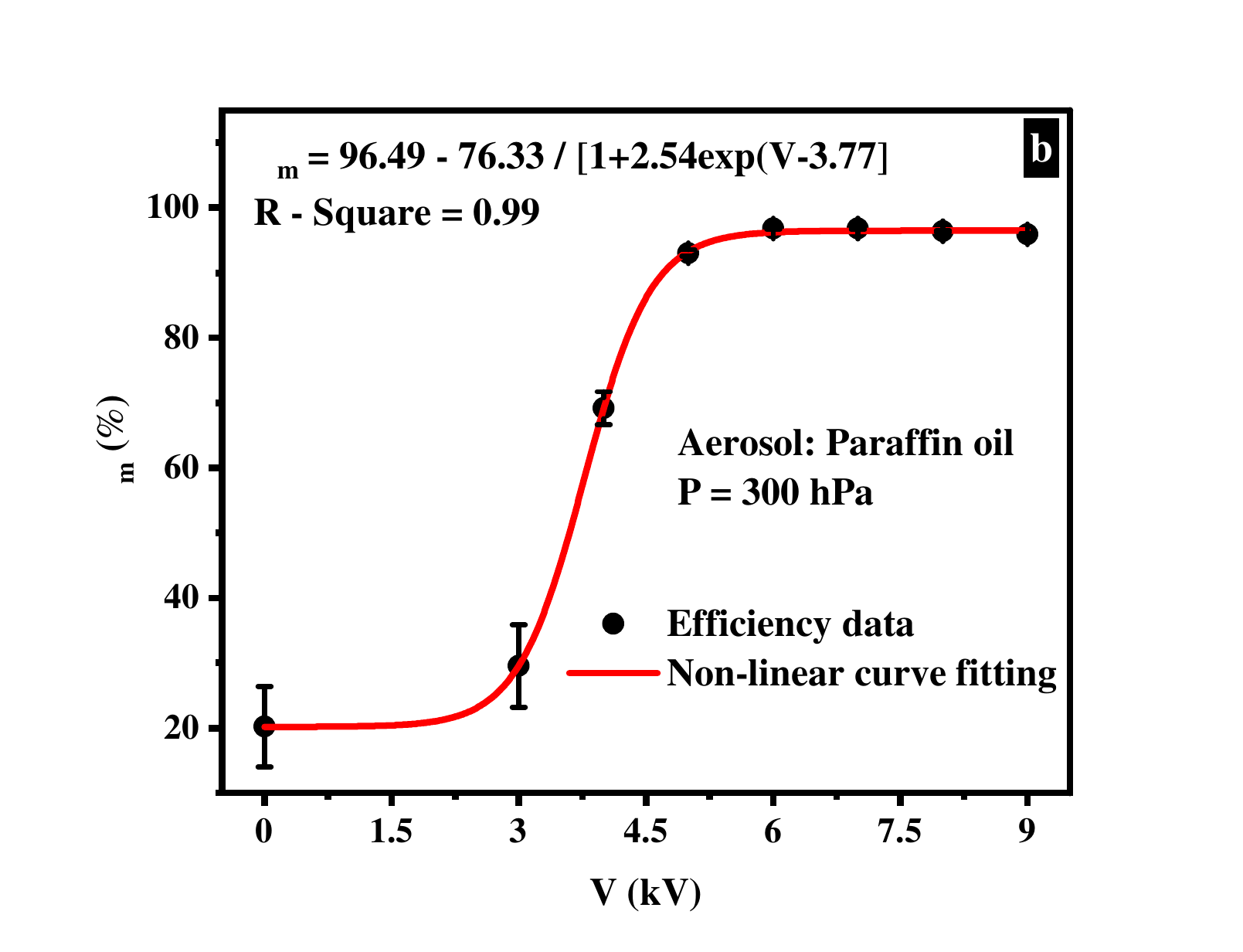}
\caption{Paraffin oil}
\label{12.}
\end{subfigure}
\caption{Scavenging efficiency on the mass-based concentration of DEHS aerosol (concentration at 300 hPa) by uncharged and charged droplets at higher potential.}
\end{figure}

\begin{figure}[h]
\begin{subfigure}[h]{0.49\textwidth}
\centering
\includegraphics[width=1.2\linewidth]{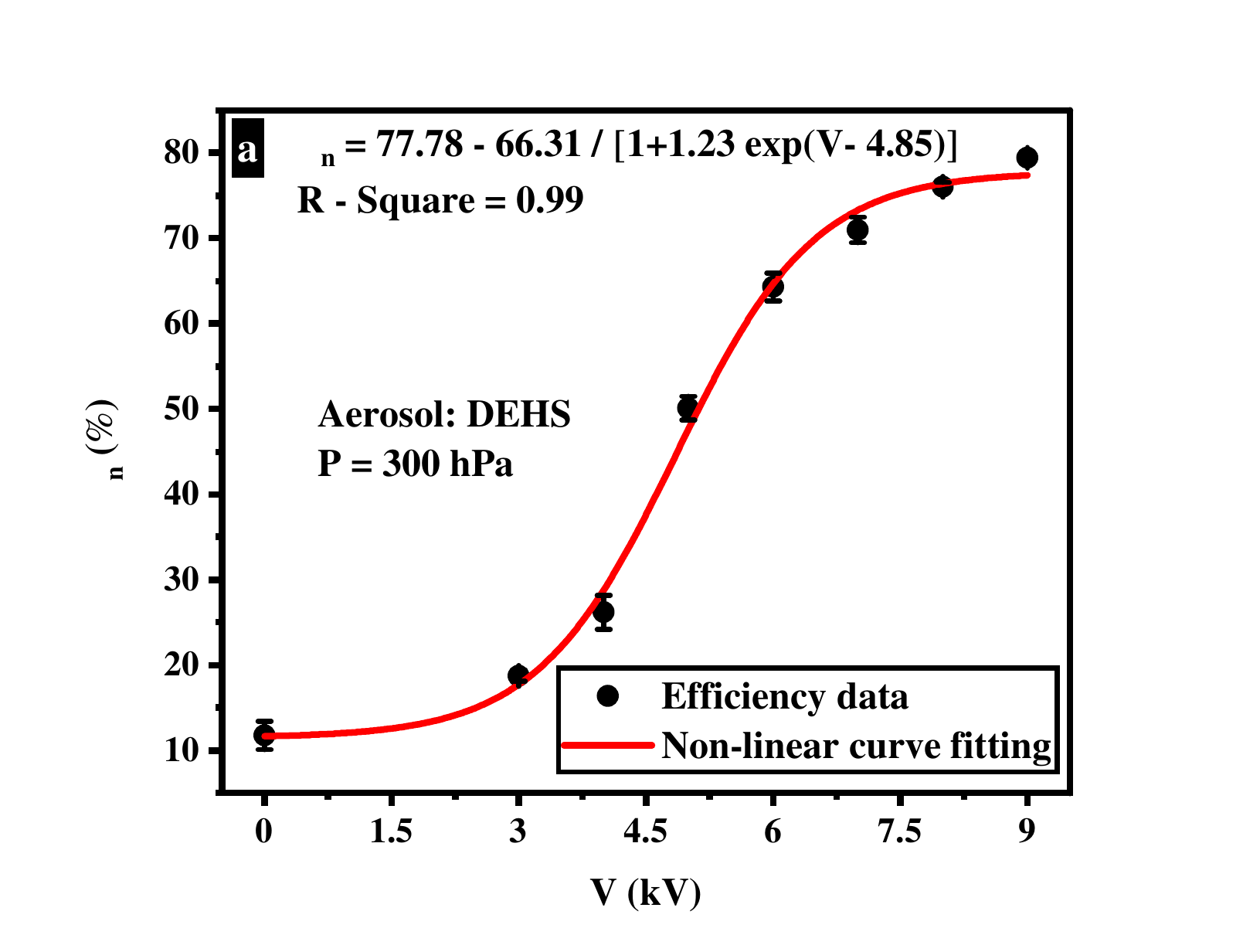}
\caption{ DEHS}
\label{13.}
\end{subfigure}
\hfill
\begin{subfigure}[h]{0.49\textwidth}
\centering
\includegraphics[width=1.2\linewidth]{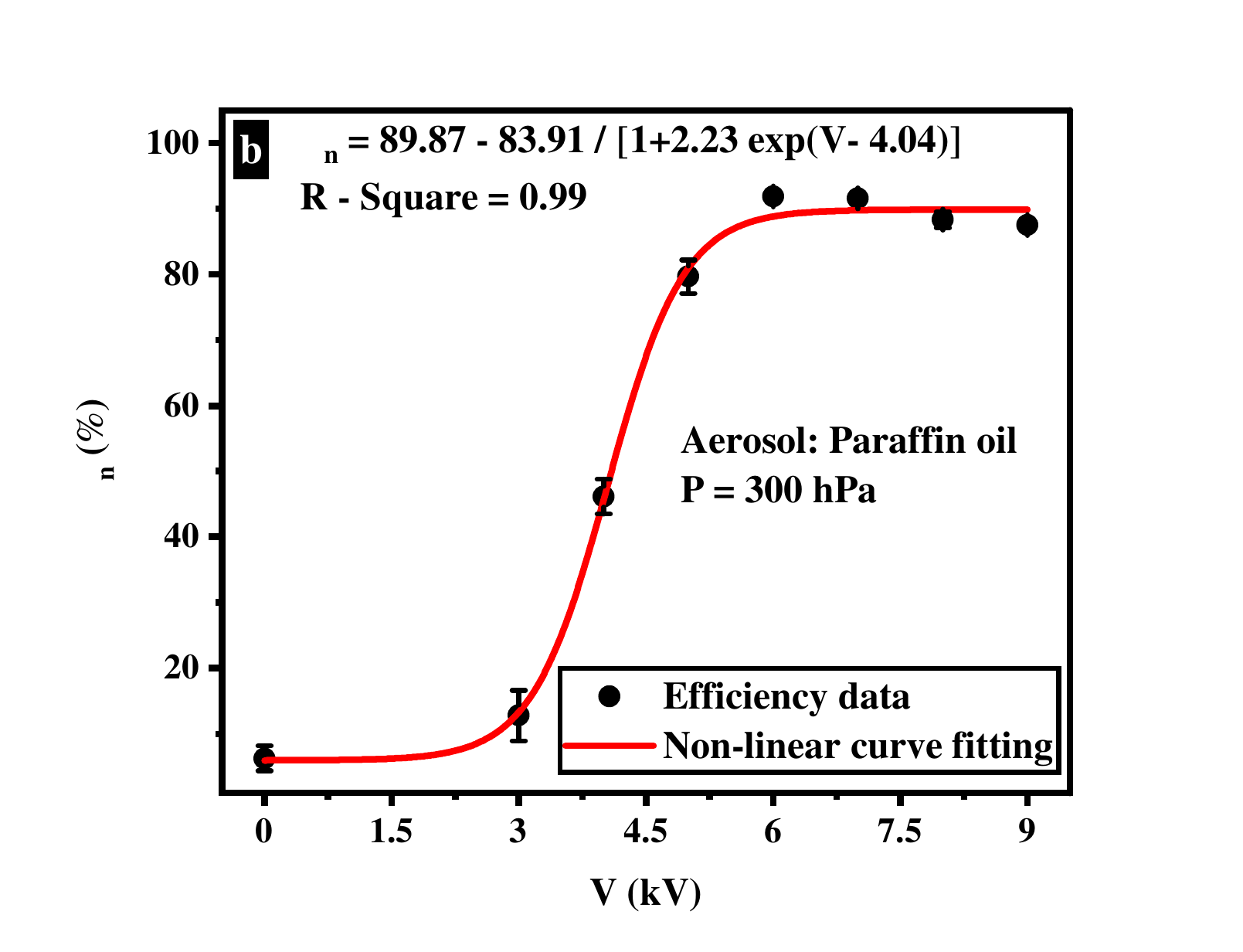}
\caption{Paraffin oil}
\label{14.}
\end{subfigure}
\caption{Scavenging efficiency on the number-based concentration of DEHS aerosol (concentration at 300 hPa) by uncharged and charged droplets at higher potential.}
\end{figure}

In the initial case of this work, the dependency of the size of the aerosol on scavenging efficiency is studied. It is observed that uncharged droplets only remove aerosols whose size is larger than 500 nm as illustrated in Figure \ref{diameter_vs_efficiency} (a). The scavenging efficiency of aerosols by uncharged droplets is determined for the two ranges of aerosol size. Here we consider the aerosol whose size lies between 500 - 700 nm as shown in Figure \ref{diameter_vs_efficiency} (a) and 700 - 1050 nm as shown in Figure \ref{diameter_vs_efficiency} (b). The uncharged droplets remove only 4\% of 500 nm size aerosol and 50\% for 700 nm size aerosol as shown in Figure \ref{diameter_vs_efficiency} (a). It is observed that with the increase of aerosol size from 500 to 700 nm, the scavenging efficiency linearly increases as shown in Figure \ref{diameter_vs_efficiency} (a). Whereas, when the size of aerosols increases from 700 to 1050 nm, the scavenging efficiency varies polynomial with order 2 as shown in Figure \ref{diameter_vs_efficiency} (b). The maximum scavenging efficiency by uncharged droplets is found to be 90\% for larger-sized aerosol (1050 nm).\\
The next analysis of this study focuses on determining the scavenging efficiency of aerosol by charged droplets. Figure \ref{diameter_vs_efficiency} (c) shows that the minimum size of aerosol that can be removed by charged droplets at 9 kV is 300 nm. This signifies that the charged droplets at 9 kV fail to remove the aerosol whose size is less than 300 nm. It is observed that charged droplets at 9 kV clean 55\% of 300 nm size aerosols. With the increase of aerosol size, the scavenging efficiency is exponentially increased as shown in Figure \ref{diameter_vs_efficiency} (c). The maximum scavenging efficiency obtained by charged droplets at 9 kV is 99.8\% for 1050 nm size aerosol. Figure \ref{diameter_vs_efficiency} (d) represents that with the increase of aerosol size and applied potential, the scavenging efficiency of aerosol also increases.\\
From the above study, it is observed that a minimum applied potential of 6 kV is necessary to remove 99\% of aerosols of size larger than 500 nm by charged droplets.\\\\
 From the above discussion, it is observed that applied potential has a significant effect on the aerosol removal from the experimental chamber. Here we investigate to estimate a correlation between the scavenging efficiency and applied potential. The scavenging efficiency of the particles is plotted with various applied electric potentials. It is observed that with the enhancement of applied potential, the scavenging efficiency of DEHS and paraffin aerosol by charged droplets also increases. The higher applied potential causes droplets to acquire more charge leading to more electrostatic force of attraction between aerosol and droplets. The overall scavenging efficiency of DEHS aerosol and paraffin oil by uncharged droplets are determined to be  55\%  as shown in Figure \ref{11.} and 20\%  as shown in Figure \ref{12.} respectively. The maximum scavenging efficiency by charged droplets at 9 kV is 99\% for both the DEHS and paraffin oil as depicted in Figures \ref{11.} and \ref{12.}.\\
 Similarly, the scavenging efficiency of aerosols based on the number concentration is illustrated. It is found that overall scavenging efficiency by uncharged droplets is 15\% for DEHS solution as shown in Figure \ref{13.} and 10\% for paraffin oil as shown in Figure \ref{14.}. The maximum scavenging efficiency by charged droplets at 9 kV is determined to be 80\% for both DEHS and paraffin oil.\\
 It is also observed that at higher applied potentials i.e., 7 kV, 8 kV, and 9 kV, the efficiency curve becomes asymptotic. It means that at higher applied potentials the efficiency of aerosol is almost equal. A relation is estimated between the scavenging efficiency of the particles and applied electric potentials with statistical curve fittings.  As shown in Figures \ref{11.}, \ref{12.}, it is discovered that in cases of mass concentration, the scavenging efficiency follows an exponential relation with applied potential with 0.99 R-squared value. As demonstrated in Figures \ref{13.}, \ref{14.}, the scavenging efficiency for number concentration follows a similar trend to mass concentration with a 0.99 R-squared value.

\subsubsection{Effect of aerosol concentrations on scavenging efficiency}
\begin{figure}[t!]
\centering
\includegraphics[width=1\linewidth]{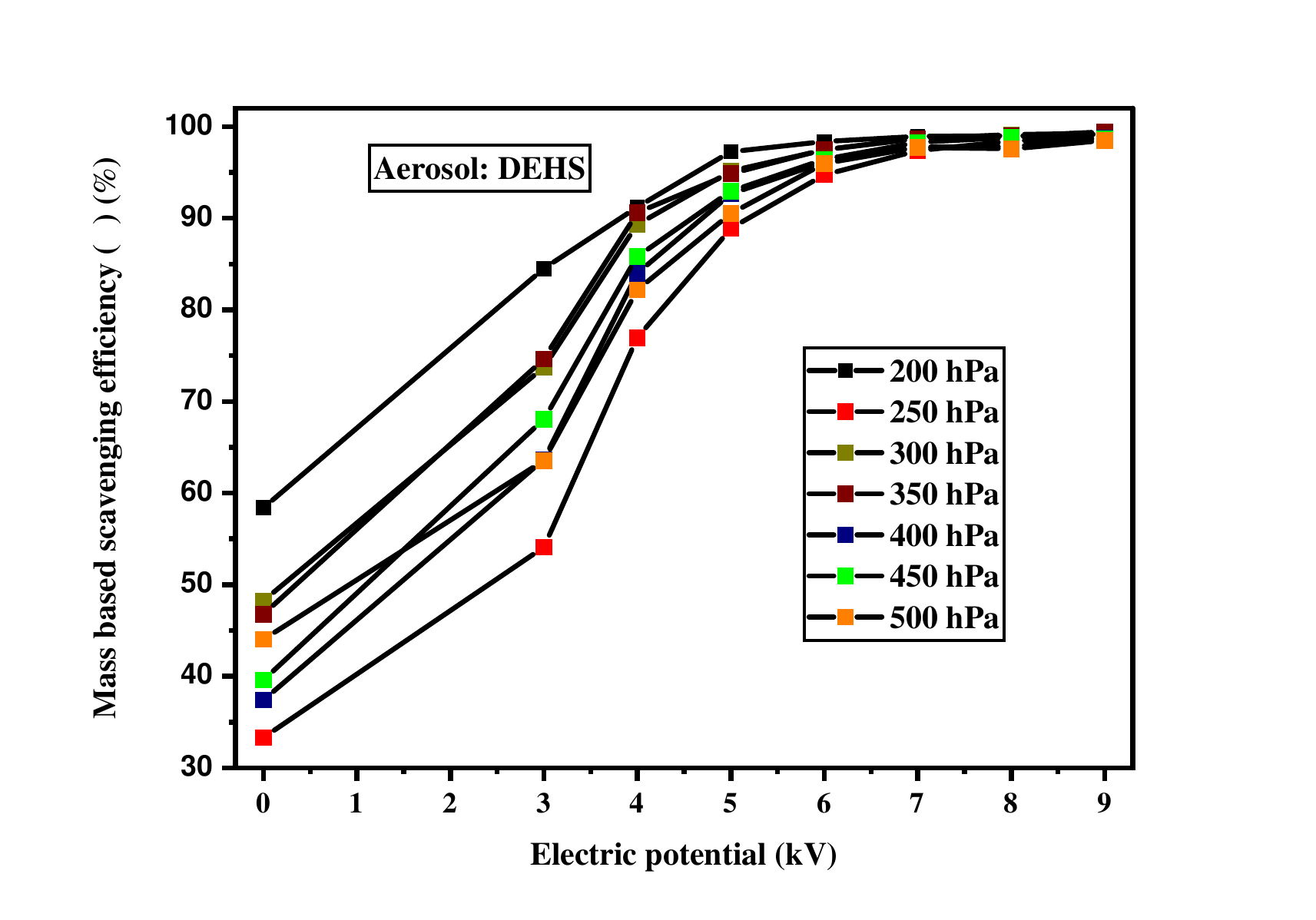}
\caption{ Scavenging efficiency on the mass-based concentration of DEHS aerosol by uncharged and charged droplets.}
\label{efficiency_all_concentrations}
\end{figure}

Since, this study is conducted by considering different concentrations (depending on the different vapor pressure of the aerosol generator i.e., 100 - 500 hPa) of aerosols, the scavenging efficiency of each concentration is estimated. It is found that the scavenging efficiency at higher electric applied potentials is very similar with all particle concentrations, as shown in Figure \ref{efficiency_all_concentrations}. This indicates that the effectiveness of charged droplets in scavenging the fine aerosols is independent of the aerosol concentration for this experimental test rig. This signifies that if the droplets acquire sufficient charge i.e., enough electrostatic force of attraction between the aerosols and particles, the charged droplets can remove 99\% of the particles, regardless of whether the concentration is high or low.
\newpage

\subsubsection{Scavenging efficiency based on mass concentration and number concentration}
\begin{figure}[h!]
\centering
\includegraphics[width=1\linewidth]{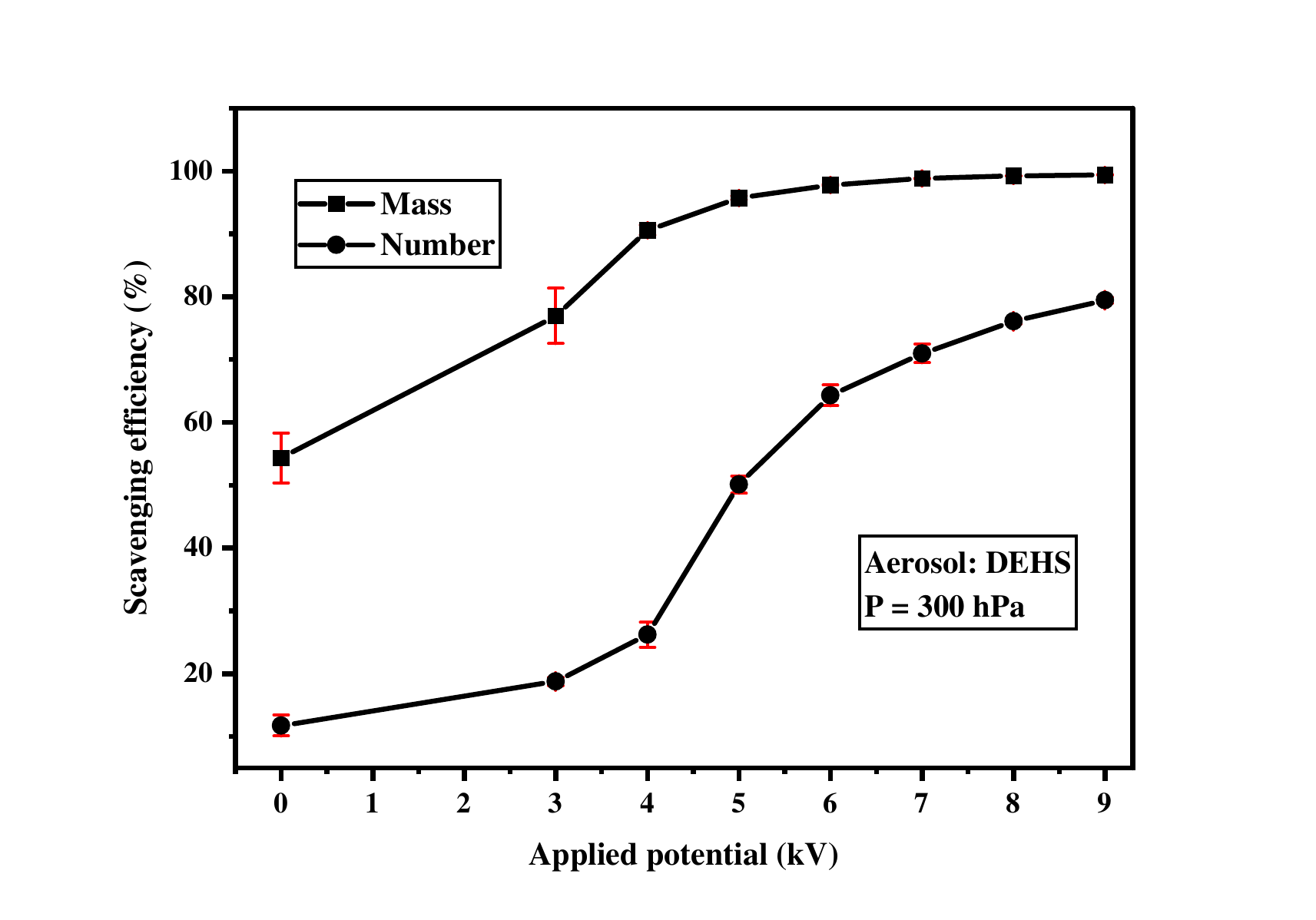}
\caption{ Scavenging efficiency comparison of DEHS aerosol (concentration at 300 hPa) by uncharged and charged droplets.}
\label{efficiency_number_vs_mass}
\end{figure}

\begin{figure}[h]
\centering
\includegraphics[width=1\linewidth]{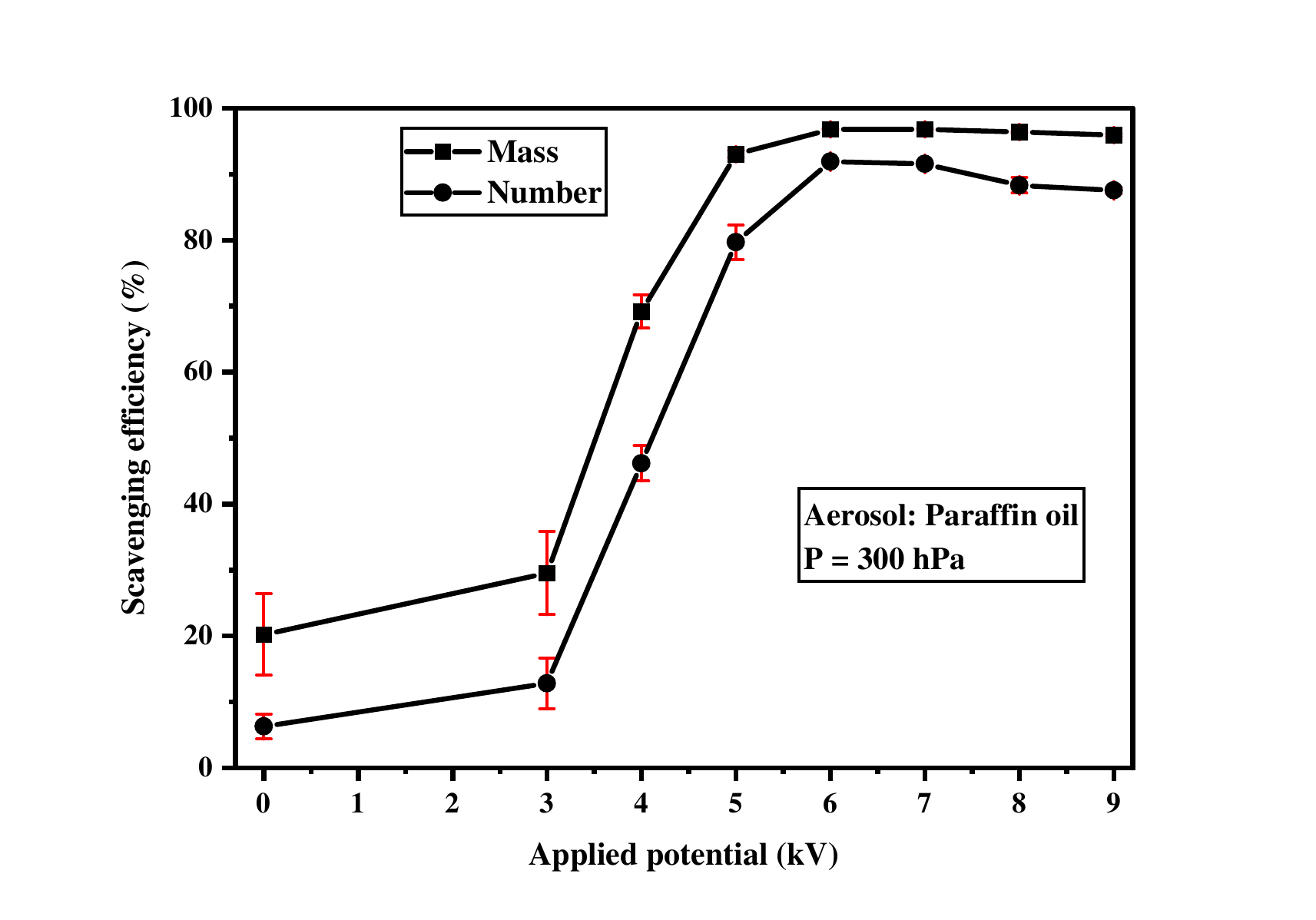}
\caption{ Scavenging efficiency comparison of paraffin oil at 300 hPa vapor pressure of atomizer by uncharged and charged droplets.}
\label{efficiency_comparison_paraffin_oil}
\end{figure}

As discussed in the above sections, all sizes of the aerosols are not removed by either uncharged or charged droplets. It tells that some sizes of the aerosols are inseparable through the wet electrostatic scrubber. It is necessary to compare the scavenging efficiency of the aerosols based on their number concentration and mass concentrations. 
Comparisons are made between the mass concentration and number concentration of the particles eliminated by uncharged and charged droplets. The results show that the scavenging efficiency of particles based on mass concentration is significantly higher than that of particles based on number concentration as shown in Figure \ref{efficiency_number_vs_mass} for DEHS aerosol and Figure \ref{efficiency_comparison_paraffin_oil} for paraffin oil. 
The maximum scavenging efficiency based on mass concentration is around 99\%, whereas it is around 80\% based on number concentration.
This is because some of the fine aerosols are inseparable through charged droplets.

\subsubsection{Scavenging efficiency comparison of paraffin oil and DEHS aerosol}
\begin{figure}[t!]
\centering
\includegraphics[width=1\linewidth]{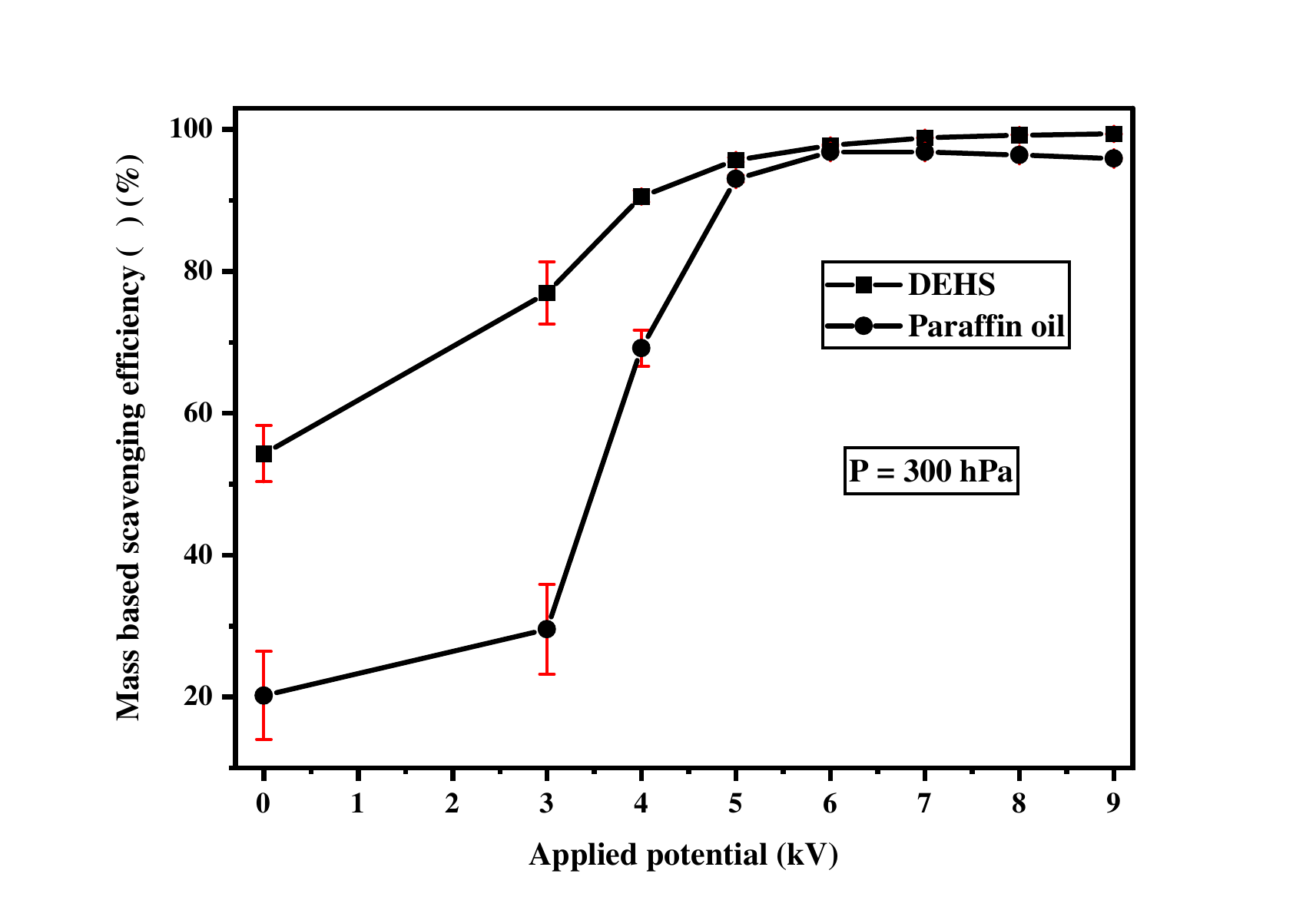}
\caption{Scavenging efficiency of two test aerosols at 300 hPa vapor pressure of atomizer by uncharged and charged droplets.}
\label{efficiency_comparison_paraffin_DEHS}
\end{figure}

The last comparison is made between the two test aerosols i.e., aerosol generated from DEHS solutions and paraffin oil. The scavenging efficiency of DEHS solution and paraffin oil is compared at a fixed concentration of aerosols generated with 300 hPa operating vapor pressure of aerosol generator. It is obtained that the DEHS solution possesses higher scavenging efficiency than paraffin oil at an applied potential between 0 to 4 kV. Paraffin oil and DEHS solution possess equal scavenging efficiency at an applied potential between 5 to 9 kV as shown in Figure. \ref{efficiency_comparison_paraffin_DEHS}. This is because of smaller particles generated from paraffin oil compared to DEHS solution. At applied potential (1 to 4 kV), impaction, interception, Brownian, and electric force both play a key role in removing the particles, whereas higher potential electric potential ( 5 - 9 kV), electrostatic force of attraction between aerosols and droplets dominates over other mechanisms.

\subsection{Theoretical discussion on particle removal mechanisms}
Based on the above study, it is observed that charged droplets demonstrate superior capture efficiency compared to uncharged droplets. To understand this hypothesis, some theoretical analysis is discussed. When the uncharged/charged droplet falls on the space occupied with particles/aerosols, the droplets interact with the particles through the following key mechanism.
\begin{itemize}
    \item Inertial impaction
    \item Direct interception
    \item Brownian diffusion
    \item Electrostatic attraction
\end{itemize}
\subsubsection{Inertial impaction}
This process is essential in eliminating particles with a diameter exceeding 5 $\mu$m \cite{jung2002analytic}. In this mechanism, particles larger in size deviate from the streamline and directly impact the surface of the droplet due to their substantial size. The efficiency of the inertial mechanism is quantified by a dimensionless parameter, the Stokes number ($stk$), as articulated in Eq.\ref{stokes no}
\begin{equation}
    stk = \frac{\rho_p^{2}d_p^2 U}{18\mu D}
    \label{stokes no}
\end{equation}
Here $\rho_p$ and $d_p$ are the density and diameter of the particle. $\mu$ is the dynamic viscosity of air. $D$ is the droplet diameter and $U$ is the relative velocity between the droplet and particle. Litch \cite{licht1988air,carotenuto2010wet} derived inertial impaction collection efficiency of the particles in wet scrubber as mentioned in Eq.\ref{impaction}.
\begin{equation}
    \eta_{imp} = (\frac{stk}{stk+0.35})^2
    \label{impaction}
\end{equation}
\subsubsection{Direct interception }
This process aids in the elimination of particles of intermediate size, ranging from 1 to 2.5 microns. In this scenario, particles adhere to the streamlined path, and because their radius exceeds the thickness of the boundary layer, they are captured when passing within one particle radius from the surface of the water droplet. The effective description of particle captures by droplets, achieved through the inertial interception mechanism, is accurately described by a non-dimensional parameter denoted as R. This parameter is defined in Eq. \ref{introduction} as the ratio of the particle diameter to the droplet diameter. 
\begin{equation}
    R = \frac{d_p}{D}
    \label{inter_parameter}
\end{equation}
Jung and Lee \cite{jung1998filtration} also established the collection efficiency of a single liquid sphere due to interception as follows:
\begin{equation}
    \eta_{int} = [(\frac{1-\alpha}{3+\sigma K})\frac{1}{D}]d_p + [(\frac{1-\alpha}{3+\sigma K})(\frac{3\sigma+4}{2D^2})]d_p^2
    \label{impaction}
\end{equation}

\subsubsection{Brownian diffusion}
Brownian diffusion emerges as the primary mechanism for collecting small particles (with sizes below 1 micron) in wet scrubbers. Given that the diffusion coefficient is inversely related to particle size, smaller particles exhibit higher diffusion coefficients. Jung and Lee \cite{jung1998filtration} formulated the subsequent expression for the diffusive collection efficiency of an individual liquid sphere, accounting for the impact of induced internal circulation within a liquid droplet:
\begin{equation}
    \eta_{diff} = 0.7\{\frac{4}{\sqrt{3}}(\frac{1-\alpha}{J+\sigma K})^{1/2}Pe^{-1/2} + 2(\frac{\sqrt{3}\pi}{4Pe})^{2/3}[\frac{(1-\alpha)(3\sigma+4)}{J+\sigma K}]^{1/3}\} 
\end{equation}
Where $\sigma$ is the density ratio of liquid to air, $\alpha$ is the packing density.
\begin{equation}
    J = 1 - \frac{6}{5}\alpha^{1/3} + \frac{1}{5}\alpha^2
\end{equation}
\begin{equation}
    K = 1 - \frac{9}{5}\alpha^{1/3} + \alpha + \frac{1}{5}\alpha^2
\end{equation}
$Pe$ is the Peclet number defined as
\begin{equation}
    Pe = \frac{DU}{D_{diff}}
\end{equation}
Where $D_{diff}$ is the diffusion coefficient of the particles defined in Eq.\ref{diffusion_coefficient} 
\begin{equation}
    D_{diff} = \frac{k_BTC_c}{3\pi\mu d_p}
    \label{diffusion_coefficient}
\end{equation}
Where $k_B$ is the Boltzmann constant, $T$ is the absolute temperature, and $C_c$ is the Cunningham slip correction factor.

\subsubsection{Electrostatic attraction}
The electrostatic plays a crucial role when the particles and droplets possess sufficient charges. The particles are attracted towards the charged droplet due to the electrostatic force of the attraction. Nielsen and Hill \cite{nielsen1976collection} developed imperial relation for obtaining the collision efficiency between the droplet particles due to the electrostatic force of attraction given in Eq.\ref{electrostatic}
\begin{equation}
    \eta_{ES} = \left\{\dfrac{15\pi}{8}(\dfrac{\epsilon_p - 1}{\epsilon_p + 2})\dfrac{4C_c[4q/(\pi a^2)]^2d_p^2}{3\pi \mu_g U\epsilon_0a}\right\}^{0.4}
    \label{electrostatic}
\end{equation}

\begin{figure}[h]
\centering
\includegraphics[width=1\linewidth]{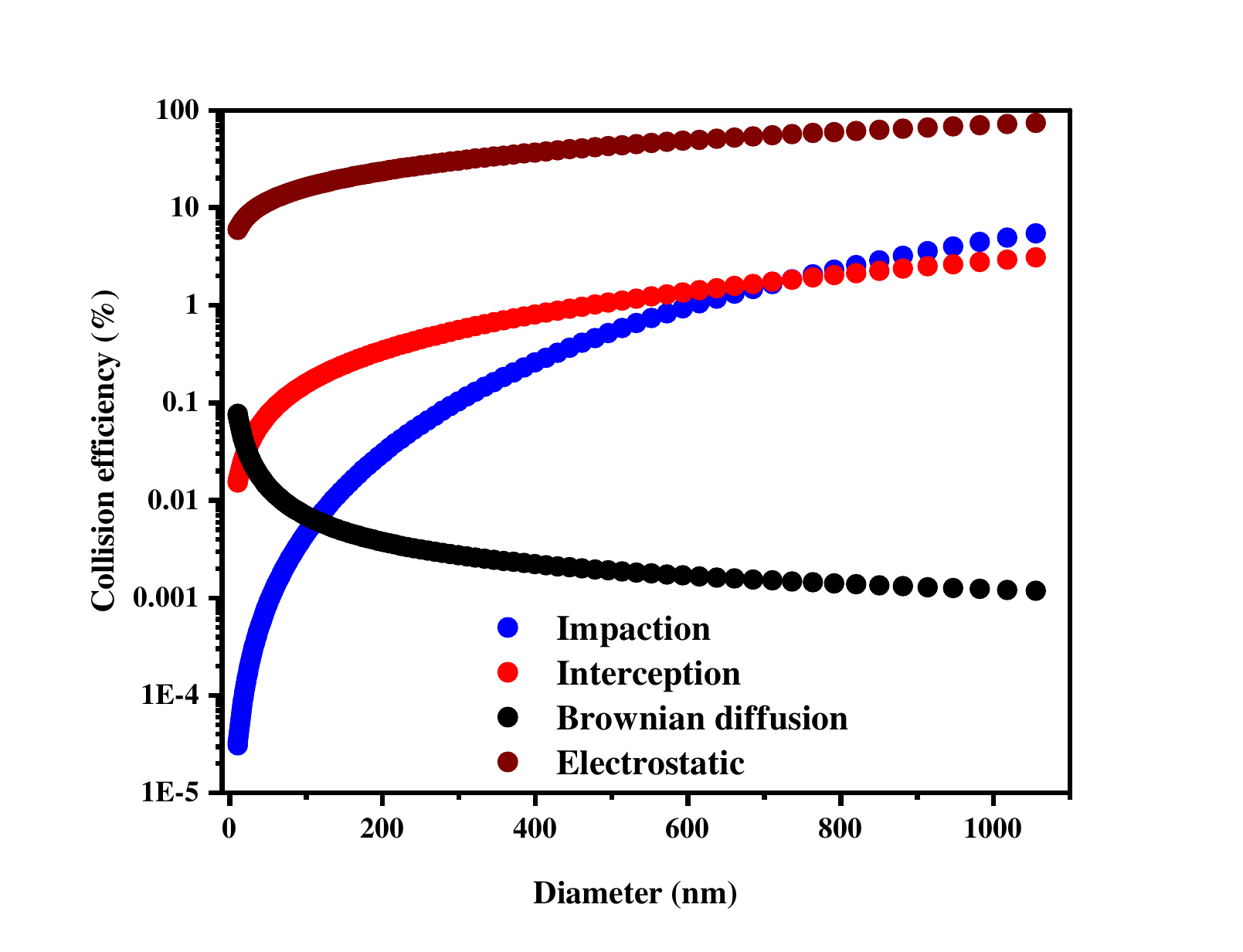}
\caption{Collision efficiency of the particle by 70-micron droplet due to inertial impaction, interception, Brownian diffusion, and electrostatic interaction at 9 kV applied potential.}
\label{all_mechanisms}
\end{figure}

 The overall collision efficiency of particles with a single charged droplet is determined using the above theoretical expressions, incorporating experimental data such as droplet size, droplet charge, the intensity of the electric field, aerosol size, and properties. Figure \ref{all_mechanisms} indicates that as the size of the particles increases, the collision efficiency through interception and inertial impaction rises. The collision efficiency of the particle due to impaction and interception is directly proportional to the square of the diameter of the particle. In contrast, the collision efficiency for Brownian diffusion is higher with smaller particles and vice versa. As the collision efficiency due to Brownian diffusion inversely depends on the size of the aerosols. Among all the mechanisms considered, electrostatic interaction predominantly contributes to the collision efficiency over other mechanisms.

\section{Conclusions} 
This study delves into the scavenging of aerosols, considering both their mass and number concentrations before and after spraying. Concerning the number concentration, a binomial distribution of particles is noted. The small distribution signifies the presence of atmospheric aerosols in the chamber before feeding particles into the chamber, along with tiny droplets generated through spray entering the Scanning Mobility Particle Sizer (SMPS). This occurrence might be attributed to the diffusion dryer's inefficiency, used to absorb water, as observed in multiple experiments. Since the SMPS counts fine water droplets as particles, the efficiency in terms of number concentration is lower compared to mass-based concentration. Therefore, it is crucial to distinguish between the efficiencies based on mass and number concentrations. From this experimental study, the following vital outcomes are observed.

\begin{enumerate}
    \item The uncharged droplets effectively remove the aerosols with a minimum size of more than 500 nm. Whereas, the charged droplets at 9 kV remove a minimum 300 nm size of the aerosols. The aerosols with 10 - 290 nm are inseparable either by uncharged or charged droplets. 
    \item The maximum scavenging efficiency of the aerosols by uncharged droplets in this experimental test rig is around 55\%. Whereas the highest collection efficiency by charged droplets is found to be 99\%.
    \item The efficacy of particle removal utilizing uncharged and charged droplets with lower potentials of +1 kV and +2 kV is approximately the same. So, eliminating the particles with an electrospray with a low potential for the same system is ineffective. The electrospray is effective for the abatement of tiny particles by charged droplets at an electric potential of at least +3 kV.
    \item With the enhancement of the applied potentials from 3 kV to 9 KV to droplets, the collection efficiency of the particles also increases. It is also observed that at higher potentials i.e., 7, 8, and 9 KV, the scavenging efficiency of the particles for all the different concentrations of aerosol is the same.
\end{enumerate}
\textbf{Acknowledgment}\\
The Authors acknowledge the Department of Science and Technology (DST/TMD/CERI/Air Pollution/2018/009), India for providing financial support for this experimental work.
\nomenclature[1]{\(stk\)}{Stoke number [-]}
\nomenclature[2]{\(\rho_p\)}{Density of particle [$kg/m^3$]}
\nomenclature[3]{\(U\)}{Relative velocity between particle and droplet [m/s]}
\nomenclature[4]{\(d_p\)}{Diameter of particle [m]}
\nomenclature[5]{\(\mu\)}{Dynamic viscosity of air [$Pa s/m^2$]}
\nomenclature[6]{\(\eta_{imp}\)}{Impaction efficiency [-]}
\nomenclature[7]{\(\eta_{int}\)}{Interception efficiency [-]}
\nomenclature[8]{\(\eta_{diff}\)}{Diffusion efficiency [-]}
\nomenclature[9]{\(\sigma\)}{Density ratio of liquid to air [-]}
\nomenclature[10]{\(\alpha\)}{Packing density [-]}
\nomenclature[11]{\(Pe\)}{Peclet number [-]}
\nomenclature[12]{\(D_{diff}\)}{Diffusion coefficient [$m^2/s$]}
\nomenclature[13]{\(k_{B}\)}{Boltzman constant [$kg m^2/s^2/K$]}
\nomenclature[14]{\(C_c\)}{Cunningham slip correction factor [-]}
\nomenclature[15]{\(T\)}{Absolute temperature [K]}
\nomenclature[16]{\(C_i\)}{Initial concentration Mass/number of aerosols without spray}
\nomenclature[17]{\(C_0\)}{Mass/number concentration of particles with uncharged spray}
\nomenclature[17]{\(C_1\)}{Mass/number concentration of particles with charged spray at 1 kV}
\nomenclature[17]{\(C_2\)}{Mass/number concentration of particles with charged spray at 2 kV}
\nomenclature[17]{\(C_3\)}{Mass/number concentration of particles with charged spray at 3 kV}
\nomenclature[17]{\(C_4\)}{Mass/number concentration of particles with charged spray at 4 kV}
\nomenclature[17]{\(C_5\)}{Mass/number concentration of particles with charged spray at 5 kV}
\nomenclature[17]{\(C_6\)}{Mass/number concentration of particles with charged spray at 6 kV}
\nomenclature[17]{\(C_7\)}{Mass/number concentration of particles with charged spray at 7 kV}
\nomenclature[17]{\(C_8\)}{Mass/number concentration of particles with charged spray at 8 kV}
\nomenclature[17]{\(C_9\)}{Mass/number concentration of particles with charged spray at 9 kV}
\nomenclature[17]{\(P\)}{Vapour pressure of aerosol generator [hPa]}
\nomenclature[17]{\(q\)}{Charge on droplet [C]}
\nomenclature[17]{\(V\)}{Applied DC potential [kV]}
\nomenclature[17]{\(E\)}{Non-uniform electric field distribution [V/m]}
\nomenclature[17]{\(E_r\)}{Non-uniform radial electric field distribution [V/m]}
\nomenclature[17]{\(E_z\)}{Non-uniform axial electric field distribution [V/m]}
\nomenclature[17]{\(a\)}{Radius of droplet [m]}
\nomenclature[17]{\(\epsilon_0\)}{Permittivity of free space [F/m]}
\nomenclature[17]{\(\eta_m\)}{Mass efficiency of the aerosol [$\%$]}
\nomenclature[17]{\(\eta_n\)}{Number efficiency of the aerosol [$\%$]}
\printnomenclature

 \bibliographystyle{elsarticle-num} 
 \bibliography{cas-refs}





\end{document}